\documentclass[lettersize,journal]{IEEEtran}
\DeclareOldFontCommand{\rm}{\normalfont\rmfamily}{\mathrm}
\DeclareOldFontCommand{\sf}{\normalfont\sffamily}{\mathsf}
\usepackage{xcolor}
\DeclareOldFontCommand{\tt}{\normalfont\ttfamily}{\mathtt}
\DeclareOldFontCommand{\bf}{\normalfont\bfseries}{\mathbf}
\DeclareOldFontCommand{\it}{\normalfont\itshape}{\mathit}
\DeclareOldFontCommand{\sl}{\normalfont\slshape}{\@nomath\sl}
\DeclareOldFontCommand{\sc}{\normalfont\scshape}{\@nomath\sc}
\usepackage{amsthm}
\usepackage{amsthm,amsmath,amssymb,lipsum}
\usepackage{mathrsfs}
\usepackage{amsmath,amsfonts}
\usepackage{algorithmic}
\usepackage{algorithm}
\usepackage{array}
\usepackage{bm}
\usepackage{tabularray}
\usepackage{graphicx}
\usepackage{float} 
\usepackage[caption=false,font=normalsize,labelfont=sf,textfont=sf]{subfig}
\usepackage{textcomp}
\usepackage{stfloats}
\usepackage{url}
\usepackage{verbatim}
\usepackage{cite}
\usepackage{epstopdf}
\usepackage{booktabs} 
\usepackage{siunitx}

\graphicspath{{figures/}} 

\newcommand{\rev}[1]{\textcolor{black}{#1}}

\begin{document}
\title{Combating Suppressive Jamming with Dynamic Agile Reconfigurable Intelligent Surface Antenna Array (DARISAA)}

\author{Yakun Ma, Qingli Yan, and Hui-Ming Wang, \emph{Senior Member, IEEE}%
	\thanks{Yakun Ma and Hui-Ming Wang are with the School of Information and Communications Engineering, Xi'an Jiaotong University, Xi'an 710049, China (e-mail: mayakun1998@163.com, xjbswhm@gmail.com).}
	\thanks{Qingli Yan is with the School of Computer Science \& Technology, Xi'an University of Posts \& Telecommunications, Xi'an 710121, China (e-mail: yql@xupt.edu.cn).}
}

\maketitle



\begin{abstract}
Suppressive jamming is a severe challenge to digital receivers in  wireless communications, since high-power jamming may cause analog-to-digital converter (ADC) overload or automatic gain control (AGC)-limited quantization blocking of weak desired signals. Analog beamforming in the spatial domain can be employed to combat suppressive jamming signals before they reach the ADC. In this paper, we introduce a comprehensive anti-jamming scheme considering both direction-of-arrival (DoA) acquisition and suppressive jamming elimination based on a Dynamic Agile Reconfigurable Intelligent Surface Antenna Array (DARISAA). DARISAA is a novel type of reconfigurable antenna array composed of many metamaterial elements, which can dynamically adapt its phase responses, enabling suppressive jamming cancellation at the antenna-end. Accurate DoA estimation schemes for jamming and desired signals are proposed based on a subspace approach by utilizing the dynamic agility of the DARISAA. By leveraging the spatial-domain preprocessing capability of DARISAA and the digital beamforming of the multi-channel receiver, an analog (antenna)-digital hybrid anti-jamming scheme is developed to effectively suppress high-power jamming and significantly improve the signal-to-interference-plus-noise ratio (SINR) even with low-resolution ADCs.  Simulation results demonstrate that the proposed comprehensive anti-jamming scheme achieves significant SINR enhancement, highlighting the advantages of DARISAA-enabled spatial front-end processing for jamming suppression.

\end{abstract}


\section{Introduction}
Wireless communication relies on the propagation of electromagnetic waves to transmit information. However, electromagnetic waves are highly susceptible to jamming from complex surrounding environments \cite{ref1}. In particular, under high-jamming conditions such as in electronic warfare, the desired signal is often overwhelmed by persistent, high-power adversary jamming. This leads to a significant degradation in signal quality and signal-to-noise ratio (SNR), ultimately compromising the  reliability and even the connectivity of wireless links \cite{ref2, ref3, ref4}.

A key technical obstacle in resisting strong electromagnetic jamming is how to address the saturation problem of analog-to-digital converters (ADCs) in the radio-frequency (RF) chains. This is particularly critical in adversarial environments where malicious attackers actively deploy high-power suppressive jamming against targeted receivers. In such environments, the jamming-to-signal ratio can reach extremely high levels (e.g., up to 90 dB), far exceeding the dynamic range of typical ADCs \cite{addref1, 13}. Without sufficient gain control or front-end suppression, when the received signal power, dominated by such overwhelming jamming, exceeds the full-scale range of the ADC, it causes significant nonlinear distortion. This nonlinearity severely degrades the fidelity of the sampled signal, complicating subsequent signal detection, recovery, and reconstruction.

\begin{figure}[!t]
	\centering
	{\includegraphics[width=0.5\linewidth]{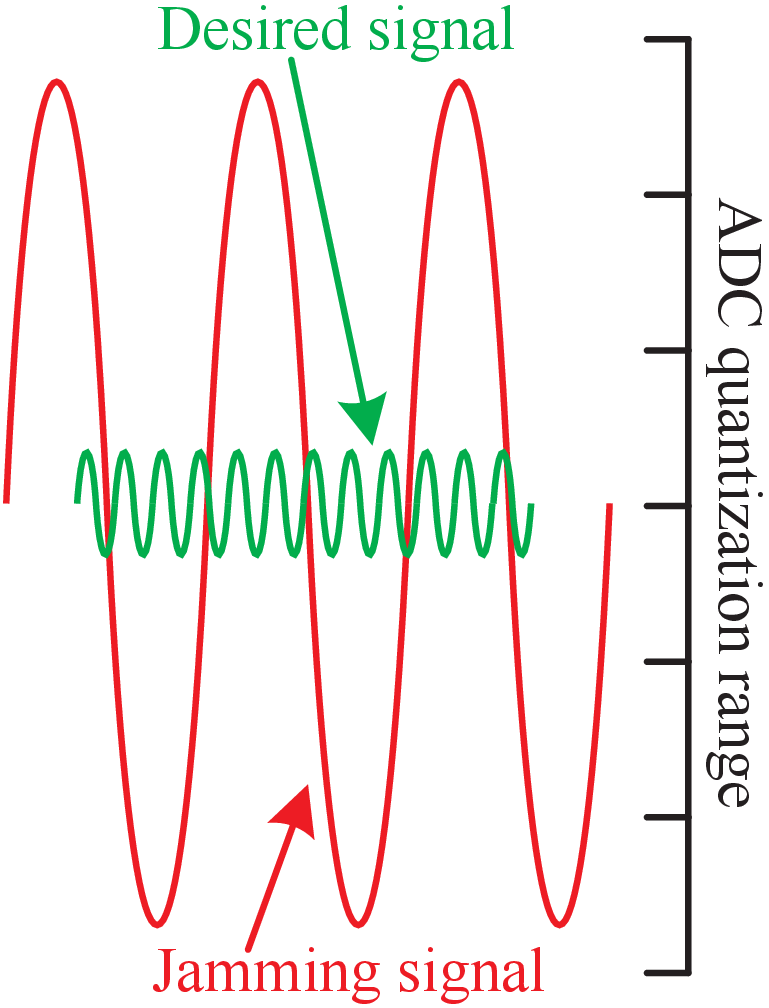}}
	\caption{Illustration of ADC blocking caused by jamming signal.}
	\label{fig1}
\end{figure}

To avoid such physical ADC overload, an automatic gain control (AGC) circuit is configured before the ADC to dynamically adjust the receive gain and maintain the ADC input signal within its valid full-scale range.
In scenarios where the jamming signal is significantly stronger than the desired signal, the AGC circuit compresses the entire received signal to prevent saturation. However, it introduces another limitation: it proportionally scales both the strong jamming and the weak desired signal. As a result, the overwhelming jamming compresses the effective dynamic range for the desired signal, wasting a large portion of the quantization range of the ADC. This significantly reduces the resolution with which the low-power desired signal can be digitized, thereby impairing the reception quality. In some cases, this compression potentially reduces the desired signal below the resolution threshold of the ADC as shown in Fig. \ref{fig1}. As a result, the desired signal may be blocked or lost during digitization.

One intuitive solution is to enhance the dynamic range by employing ADCs with more quantization bits. While this can help accommodate stronger jamming, it comes at the cost of substantially increased power consumption and hardware cost, and thus relying solely on increasing ADC resolution is not an effective strategy for combating high-power jamming. Consequently, there is an urgent need to explore new methods that can suppress the adverse impact of strong electromagnetic jamming while preserving the integrity of the weak desired signal. Developing jamming-resilient receiver architectures that work efficiently under the constraints of ADC  resolution remains a critical and open research challenge.

An effective approach for mitigating strong jamming is to attenuate it in the analog domain of the RF chain before it enters the ADCs. Specifically, applying analog filtering at the RF front-end or even at the antenna  can suppress jamming signals. This allows even low-resolution ADCs to achieve satisfactory reception performance and significantly reduce ADC power consumption.
A classical strategy involves introducing adjustable nulls in the antenna radiation pattern by manipulating the feed network. For example, in \cite{3} and \cite{4}, deep nulls are generated by injecting signals with tunable amplitude and phase into multiple antenna elements to cancel jamming in specific directions. While effective, these solutions require precise and often rigid feed configurations, which inherently restricts their flexibility and angular coverage.

To address these limitations, spatial-domain jamming cancellation methods have been proposed. In \cite{5}, a uniform linear array (ULA) with $1\times 8$ elements achieves a 20 dB null depth through spatial cancellation. However, this approach compromises the amplitude and phase coherence of the output, which can adversely impact downstream signal processing. Similarly, the authors in \cite{6} introduce a multi-channel front-end circuit with amplitude and phase weighting, enabling spatial coupling-based null generation.  Furthermore, spatial-domain suppression has also been extended to multiple-input multiple-output (MIMO) systems. The spatial matrix filtering method in \cite{7} achieves directional jamming rejection by dividing the angular domain into passband and stopband regions. Likewise, an adaptive null-domain filtering method proposed in \cite{8} reduces ADC resolution by 4 bits and achieves approximately 90$\%$ power savings at 2 bits/s/Hz spectral efficiency. While these methods are theoretically effective, they often suffer from high hardware complexity due to reliance on intricate analog circuits or require accurate prior knowledge of jamming directions.

Hybrid analog-digital beamforming architectures have emerged as a promising solution, aiming to overcome the hardware cost and power consumption bottlenecks imposed by massive RF chains and high-resolution ADCs. Building upon this, several works \cite{9,10,11} explore hybrid analog-digital beamforming to mitigate jamming prior to quantization. These works demonstrate improved energy efficiency and robustness, validated by over-the-air experiments and optimization of ADC resolution. Nevertheless, most of these approaches assume perfect channel state information (CSI) and require multiple high-resolution RF chains. Such requirements remain impractical in many real-world deployment scenarios, especially when no prior information about the desired or jamming signals is available. Along these lines, the work in \cite{ref5} integrates jamming direction-of-arrival (DoA) estimation into the design of an optimization-based solver to design a phase-only analog jamming-nullifying beamformer (AINB). However, its accompanying direction-finding algorithm suffers from an inherent inability to resolve coherent multipath signals.

These challenges highlight a critical trade-off between system flexibility, jamming suppression capability, and hardware cost. Within this context, we propose a new solution based on the Dynamic Agile Reconfigurable Intelligent Surface Antenna (DARISA) system.

DARISA is a novel type of reconfigurable antenna composed of many metamaterial elements, which is first proposed in \cite{ref11}. DARISA is capable of dynamically modifying its antenna radiation response via controlling the  amplitude and phase response of each metamaterial element, thereby achieving reconfigurability \cite{ref11}. 
Compared with traditional antennas, the reconfigurable nature of DARISA allows it to exhibit differentiated reception characteristics for incident signals from various directions. This unique property endows it with the potential to emulate analog-domain spatial filtering functions of conventional antenna arrays. Firstly, the rapid agility of DARISA enables it to sample the same transmitted baseband symbol multiple times using various analog configurations within a single symbol period, which facilitates fine-grained detection of multipath information of both jamming and desired signals. This characteristic is highly beneficial for the suppression of jamming in subsequent signal processing. Secondly, each metamaterial element of DARISA is independently tunable, affording high flexibility in controlling the spatial response. Lastly, multiple DARISAs form a DARISA array (DARISAA), each connected to an RF chain. DARISAA enables jamming suppression in the RF analog front-end (antenna-end), which significantly reduces system cost and hardware complexity.

\subsection{Contributions and Paper Organization}
The main contributions of this paper are summarized as follows:
\begin{itemize}
	\item[$\bullet$]
A DARISAA-assisted anti-jamming receiver architecture is proposed for wireless communication systems operating under suppressive jamming and limited-resolution ADC constraints. By leveraging the reconfigurable characteristics of DARISAA, directional suppression of high-power jamming is performed in the analog spatial domain (antenna domain) before ADC quantization, effectively relaxing the dynamic range and resolution requirements of ADCs.
\item[$\bullet$]
A subspace-based DoA estimation approach utilizing the temporal agility of DARISAA response is proposed. By performing rapid time-domain reconfiguration of the DARISAA, the proposed method achieves temporal smoothing of the received covariance matrix, which enables accurate estimation of the incident angles of coherent jamming and desired multipath signals without sacrificing the spatial resolution of DARISAA.
\item[$\bullet$]
A comprehensive anti-jamming processing pipeline is constructed, consisting of four stages: (i) Coherent multipath DoA estimation of the jamming using DARISAA; (ii) Directional suppression of suppressive jamming at the DARISAA; (iii) coherent multipath DoA estimation of the desired signal; and (iv) hybrid analog (DARISAA end)-digital beamforming for suppressive jamming suppression and desired signal enhancement. This enables a spatially adaptive anti-jamming strategy that fully exploits DARISAA’s configurability.
\item[$\bullet$]
Comprehensive simulation results validate the effectiveness of the proposed architecture and algorithm. Even with low-resolution ADC, the system achieves jamming suppression and accurate signal recovery. The spatial pre-processing capability provided by DARISAA significantly enhances the signal-to-interference-plus-noise ratio (SINR) and reduces overall power consumption, offering a promising solution for future low-power, jamming-resilient wireless receivers.
\end{itemize}

The rest of this paper is organized as follows. In Section \ref{2}, we present the DARISA architecture and the system model for the jamming suppression. Section \ref{3} details the coherent jamming DoA estimation using DARISAA. Section \ref{4} then describes the DARISAA jamming suppression, enabling robust DoA estimation for coherent desired signals. Building on these estimates, Section \ref{5} proposes the hybrid analog (DARISAA end)-digital beamforming to maximize SINR. The computational complexity is analyzed in Section \ref{6}, followed by extensive performance validation in Section \ref{7}. Section \ref{8} concludes the paper.

\subsection{Notations}
Throughout this paper, $a$ (or $A$), $\bf{a}$ and $\bf{A}$ stand for a
scalar, a column vector and a matrix, respectively. The transpose, conjugate transpose, and complex conjugate
are denoted by ${\left(  \cdot  \right)^T}$, ${\left(  \cdot  \right)^H}$ and ${\left(  \cdot  \right)^ * }$, respectively. 
$\left|  \cdot  \right|$, ${\rm{Tr}}( \cdot )$ and ${( \cdot )^{ - 1}}$ denote the determinant (or modulus for a complex variable),  trace and inverse of a matrix.
$ \odot $ is the Hadamard product of two matrices. 
${\mathbb{C}^{M \times N}}$ is a complex space with $M \times N$ dimensions.
$\mathbb{E}\left[  \cdot  \right]$ denotes
the expectation operator.
${{\bf{I}}_N}$ denotes the $N \times N$ identity matrix and ${{\bf{1}}_{N \times M}}$ denotes an $N \times M$ all-one matrix, respectively.
${\mathop{\rm Re}\nolimits} ( \cdot )$ represents the real part.
${\cal C}{\cal N}\left( {b,\sigma _{}^2} \right)$ represents
the complex normal distribution with mean $b$ and variance ${\sigma _{}^2}$.
Finally, ${\rm{blkdiag}}({{\bf{A}}_1},{{\bf{A}}_2},\ldots,{{\bf{A}}_N})$ denotes a block diagonal matrix with the matrices ${{\bf{A}}_1},{{\bf{A}}_2},\ldots,{{\bf{A}}_N}$ as its diagonal blocks. Similarly, $\rm{diag}(\bf{A})$ creates a diagonal matrix from the diagonal elements of $\bf{A}$.

\section{System Model and Problem Formulation}\label{2}

\subsection{DARISA Architecture}
\begin{figure}[!t]
	\centering
	{\includegraphics[width=0.85\linewidth]{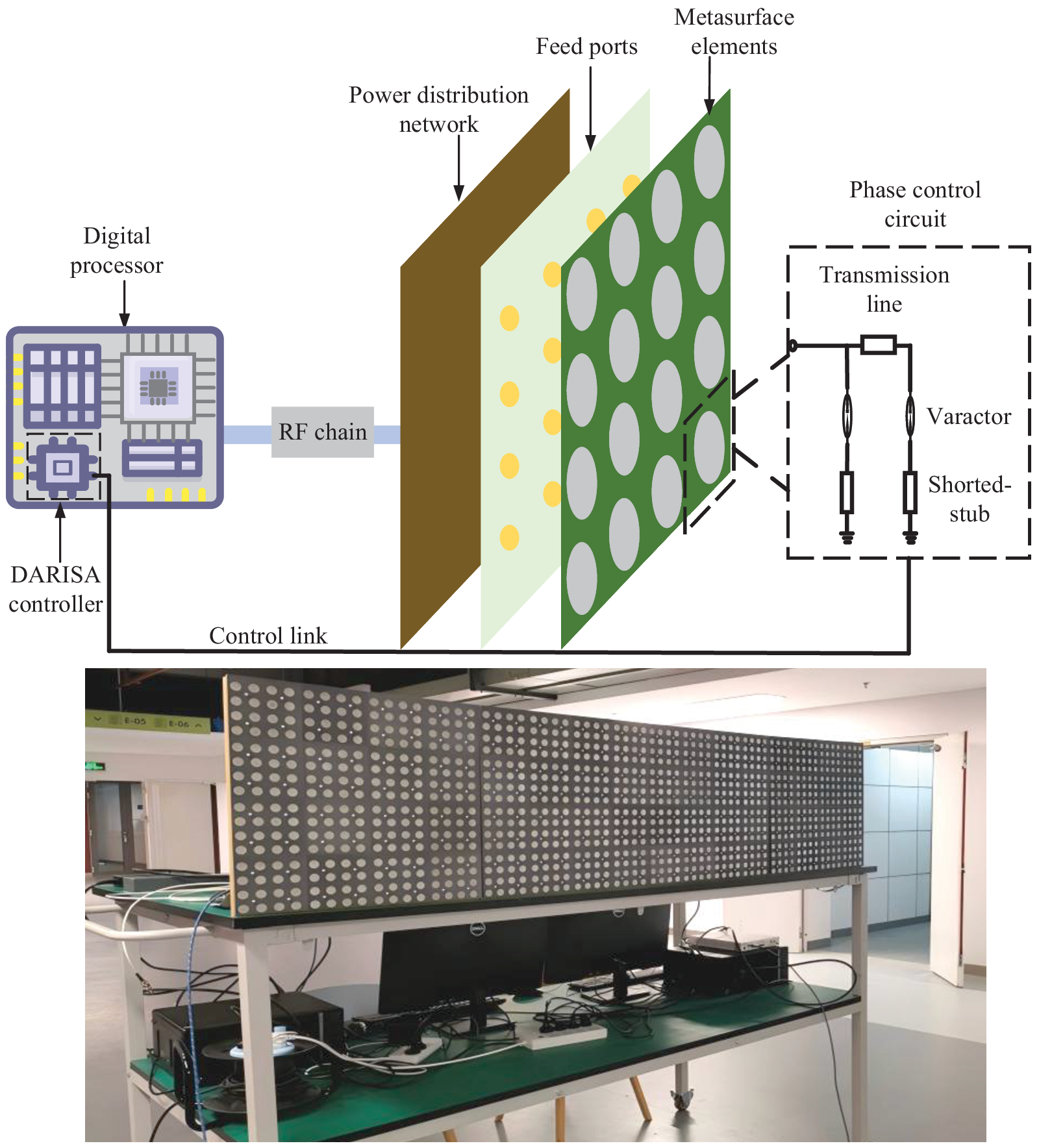}}
	\caption{DARISA architecture and DARISAA communication system prototype.}
	\label{fig_1}
\end{figure}

We first give a brief introduction of our DARISA architecture, which has been  described in \cite{ref11} in detail.
The architecture of a DARISA is depicted in Fig. \ref{fig_1}, which consists of numerous metamaterial elements with corresponding phase control circuits, a parallel feed network, and feed ports. The phase control circuit of each element is responsible for adjusting the phase responses of the metamaterial element, which is composed of varactors, transmission lines, and short stubs. Specifically, by regulating the DC voltage of the phase control circuits, the varactors can be tuned to adjust the phase response continuously. Although this hardware architecture has been applied to multi-user communications \cite{ref12, addref12} and sensing \cite{ref13}, a key characteristic of DARISA is its dynamic agility, enabled by the nanosecond-level response time of the varactor diodes, as detailed in  \cite{ref11}. This capability allows the control circuit to reconfigure the metasurface phase response with high agility, enabling multiple alterations of DARISA spatial response patterns within a single symbol period. The parallel feed network facilitates the electromagnetic interface between the feed ports and the metamaterial elements. Based on the reciprocity of passive components, this network can function as a power distribution network in transmission mode or as a signal combining network in receiving mode. These four parts are tightly connected, which ensures a compact antenna architecture while facilitating easy expansion of the size of the antenna array. Although the DARISA hardware supports dual-mode operation, this paper focuses only on its implementation as a receiver to address the suppressive jamming challenge. When it is used as a receive antenna, all metamaterial elements are excited simultaneously by the impinging signal and fed to the RF chain  through the parallel feed network and sequentially to the digital processor.

We then consider a DARISAA-based MIMO receiver equipped with multiple DARISAs. Each DARISA is connected to the ADC through the RF chain. The digital signal processing is conducted to deal with the multiple-channel received signals, such as DoA estimation, jamming cancellation and signal demodulation. 
\rev{The DARISAA receiver considered in this work is fundamentally different from both passive and active intelligent reflecting surface (IRS) architectures. Passive IRSs are external reflecting surfaces without RF/ADC connections, while active IRSs introduce reflection-type amplification to compensate for double path loss and improve coverage or interference management \cite{Wu2025, DongWang2022}. However, they still act as reflecting nodes in the propagation environment rather than receiver antenna front-ends. In suppressive jamming scenarios, the high-power jamming often arrives at the receiver through a direct or dominant path and occupies the AGC/ADC dynamic range. Such receiver-side ADC blocking is difficult to directly mitigate by an external reflecting IRS, even an active one, unless highly accurate coherent cancellation of the direct jamming path is realized. In contrast, DARISAA is directly connected to RF chains as a reconfigurable receive antenna array, enabling analog-domain spatial nulling before AGC and ADC quantization.}

\subsection{Anti-jamming System Model}
\begin{figure}[!t]
	\centering
	\includegraphics[width=0.8\linewidth]{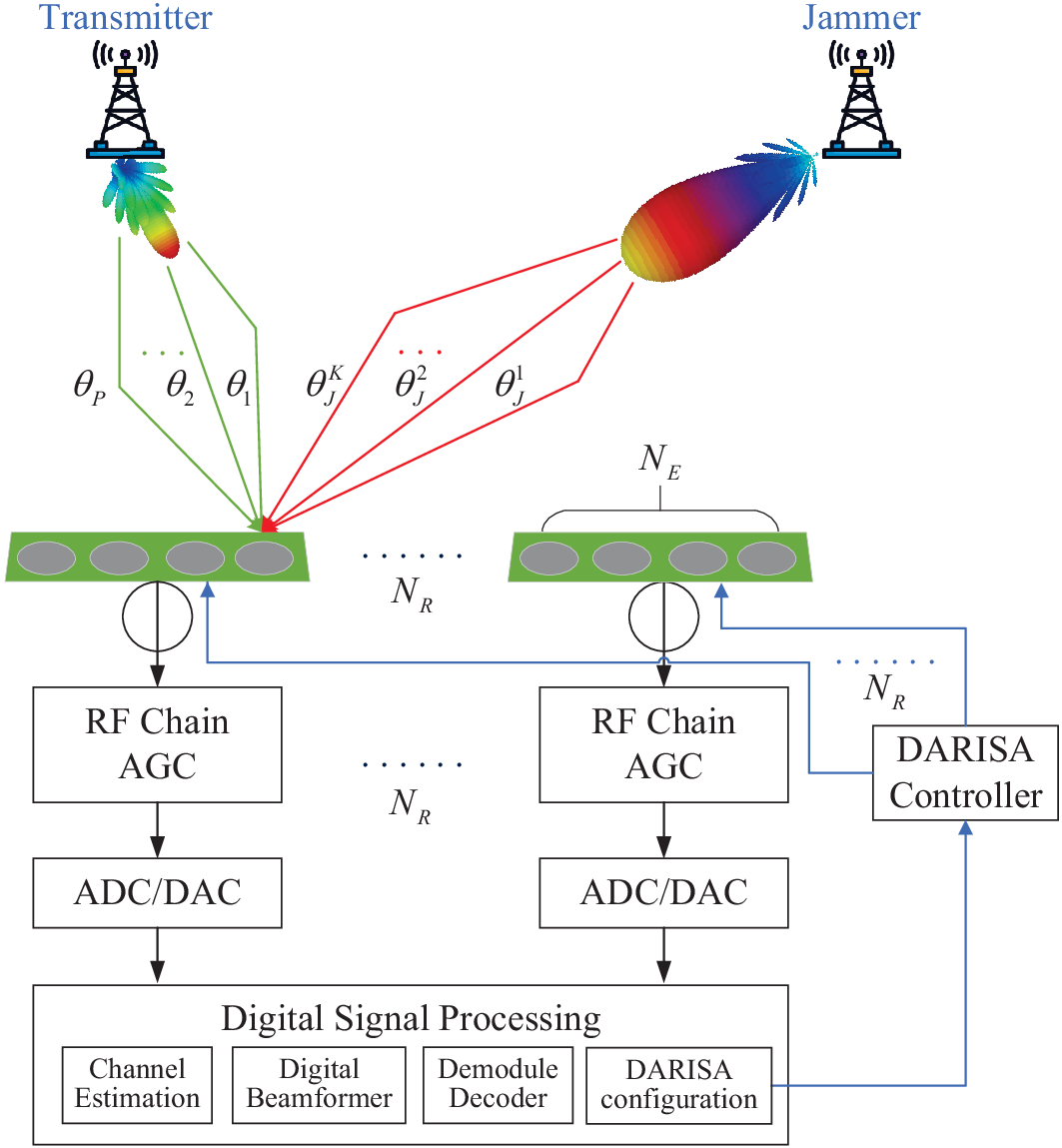}
	\caption{System model for a DARISAA anti-jamming receiver.}
	\label{fig_2}
\end{figure}

Figure \ref{fig_2} depicts the anti-jamming receiving model based on a DARISAA. The system comprises a single-antenna transmitter, a DARISAA-enabled receiver, and a single-antenna adversarial jammer. In the presence of adversary jamming, the jammer emits high-power jamming signals intended to overwhelm the desired low-power signals from the transmitter. 
In the AGC-equipped receiver considered in this paper, high-power jamming may dominate the AGC setting and compress the weak desired signal below the effective quantization resolution of the ADC, resulting in ADC blocking.
The DARISAA receiver is able to combat such distortion by dynamically adjusting the phase shifts of the DARISA elements to eliminate jamming via a specially designed DARISAA response. 

For convenience, the DARISAA considered  in the receiver is a ULA with ${N_R}$ DARISAs and corresponding RF chains. We assume the DARISAA comprises a total of $N$ adjustable metasurface elements, such that each DARISA consists of ${N_E} = N/N_R$ elements, each of which can independently phase-modulate the signal. The desired and jamming signals are ${s} \in {\mathbb{C}^{1 \times 1}}$ and $s_J \in {\mathbb{C}^{1 \times 1}}$, respectively. Due to the multipath effect, the desired signal $s$ arrives at the receiver from $P$ different directions ${\theta_1},{\theta_2},\ldots, {\theta _P}$, and the jamming signal $s_J$ arrives at the receiver from $K$ different directions $\theta _J^1,\theta _J^2,\ldots,\theta _J^K$. The antenna response vector of the ${n_r}$-th DARISA is defined as ${\bf{v}}_{n_r} \in {\mathbb{C}^{{N_E} \times 1}}$ and
\begin{equation}
\label{deqn_ex1a}
	{\bf{v}}_{{n_r}} = 1/\sqrt {{N_E}}{\left[ {\mathbf{v}_{n_r}(1),\mathbf{v}_{n_r}(2),...,\mathbf{v}_{n_r}(N_E)} \right]^T},
\end{equation}
where $\mathbf{v}_{n_r}({n_e})$ is the response of the ${n_e}$-th metamaterial element of the ${n_r}$-th RF chain. It can be expressed as $\mathbf{v}_{n_r}({n_e})=  {e^{ - j\alpha _{{n_r},{n_e}}}}$ and ${\alpha _{{n_r},{n_e}}}$ is the associated phase shift. ${\mathbf{v}}_{{n_r}}$ could be considered as the antenna response pattern of each DARISA.
\footnote{\rev{Although the DARISA hardware can in principle tune both amplitude and phase responses of each element, this work adopts phase-only control. In receive mode, the elements are equal-gain combined, and thus their amplitudes are set identically and only the phases are optimized. In addition, this phase-only setting enables a fair comparison with conventional hybrid analog-digital antenna arrays.}}

In this case, the received signal  ${\bf{x}} \in {\mathbb{C}^{{N_R} \times 1}}$ after passing through the DARISAA and prior to reaching the ADC can be expressed as
\begin{equation}
	\label{eq1}
	{\bf{x}} = {{\bf{V}}^H}\left( {{\bf{A}\bm{\beta }}s + {{\bf{A}}_J}{{\bm{\beta}}_J}{s_J}{\bf{ + n}}} \right),
\end{equation}
where ${\bf{A}} = \left[ {{\bf{a}}\left( {{\theta _1}} \right),{\bf{a}}\left( {{\theta _2}} \right),\ldots, {\bf{a}}\left( {{\theta _P}} \right)} \right]$ and ${{\bf{A}}_J} = \left[ {{\bf{a}}\left( {\theta _J^1} \right),{\bf{a}}\left( {\theta _J^2} \right),\ldots,{\bf{a}}\left( {\theta _J^K} \right)} \right]$ represent the array manifold matrices of the desired and jamming signals, respectively, where ${\bf{a}}\left( {{\theta _i}} \right)$ is in the form of ${\bf{a}}\left( {{\theta _i}} \right) = {\left( {1,{e^{ - j2\pi d/\lambda \sin {\theta _i}}},\ldots,{e^{ - j\left( {N - 1} \right)2\pi d/\lambda \sin {\theta _i}}}} \right)^T}$ and ${\bf{a}}\left( {\theta _J^k} \right), \forall k$ has the similar form.  ${\bm{\beta }} = {\left[ {{\beta _1},{\beta _2},\ldots,{\beta _P}} \right]^T}$ and ${{\bm{\beta }}_J} = {\left[ {\beta _J^1,\beta _J^2,\ldots,\beta _J^K} \right]^T}$  represent the complex attenuation coefficients vectors of the multipath of the desired and  jamming signals, respectively. ${\bf{V}} \in {\mathbb{C}^{N \times {N_R}}}$  represents the analog phase shifting matrix of the DARISAA, which
is a block diagonal matrix that can be expressed as
\begin{equation}
	\label{eq2}
{{\bf{V}}} = {\rm{blkdiag}}({{\bf{v}}_1},{{\bf{v}}_2},\ldots,{{\bf{v}}_{{N_R}}}),
\end{equation}
where ${\bf{v}}_{n_r} \in {\mathbb{C}^{{N_E} \times 1}}, n_r = 1,2, \cdots, {N_R}$. ${\bf{n}} \in {\mathbb{C}^{N \times 1}}$ is additive white Gaussian noise with ${\bf{n}} \sim {\cal C}{\cal N}\left( {{\bf{0}},\sigma_n^2{\bf{I}}_N} \right)$.

\begin{figure}[!t]
	\centering
	{\includegraphics[width=0.9\linewidth]{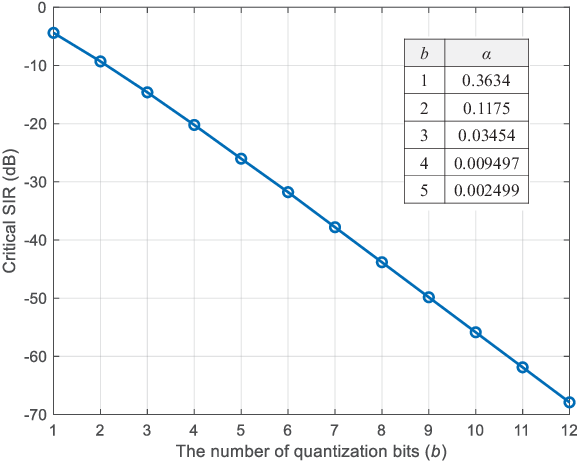}}
	\caption{Critical SIR versus the number of quantization bits.}
	\label{fig_adc}
\end{figure}

When a signal ${\bf{x}}$ passes further through the ADCs, considering the quantization noise, the received signal ${\bf{z}} \in {\mathbb{C}^{{N_R} \times 1}}$ at the digital processor can be expressed as \cite{12}
\begin{equation}
	\label{eq3}
	{\bf{z}} = \left( {1 - \alpha } \right){\bf{x}} + {{\bf{n}}_q},
\end{equation}
where ${{\bf{n}}_q}$ is the additive Gaussian quantization noise with zero mean and covariance matrix as \cite{13, 14}
\begin{equation}
	\label{eq4}
	{\bf{R}}_{n_q} = \alpha \left( {1 - \alpha } \right){\rm{diag}}\left( {{{\bf{R}}_x}} \right),
\end{equation}
where $\mathbf{R}_x = \mathbb{E}[\mathbf{x}\mathbf{x}^H]$ denotes the covariance matrix of the analog signals at the input of the ADCs. The selection of $\alpha $ is related to the number of quantization bits $b$ in the ADC. For a nonuniform scalar minimum mean square error quantizer of Gaussian random variables, for $b \le {\rm{5}}$, the values of $\alpha $ are presented in Fig. \ref{fig_adc}. \rev{For $b > 5$, $\alpha$ can be approximated as $\alpha  = \frac{{\pi \sqrt 3 }}{2}{2^{ - 2b}}$ \cite{13, GershoGray1992}.}
Fig. \ref{fig_adc} illustrates the relationship between the critical signal-to-interference ratio (SIR) and the number of quantization bits for a detectable desired signal. When the SIR is greater than this critical value, the desired signal can pass through the ADC, otherwise, it will be blocked by the ADC.

Finally, the signal ${\bf{z}}$  passes through the digital anti-jamming beamformer ${\bf{w}}\in {\mathbb{C}^{{N_R} \times 1}}$ and the output is $y$, namely,
\begin{equation}
	\label{eq5}
	\begin{split}
		y & =  {{\bf{w}}^H}{\bf{z}}\\
		& = {{\bf{w}}^H}\left( {\left( {1 - \alpha } \right){{\bf{V}}^H}\left( {{\bf{A}\bm{\beta }}s + {{\bf{A}}_J}{{\bm{\beta }}_J}{{{s}}_J}{\bf{ + n}}} \right) + {{\bf{n}}_q}} \right). 
	\end{split}
\end{equation}

{\emph {Remark 1:}}
\rev{The linear additive quantization noise model in Eq. (\ref{eq3}) is used under a non-overload assumption enabled by the AGC. Specifically, the AGC adjusts the receive gain such that the ADC input remains within the full-scale range, and hence hard clipping or nonlinear ADC saturation is avoided. The ADC blocking considered in this paper refers to an AGC-limited quantization effect rather than physical clipping: high-power jamming dominates the AGC setting and compresses the weak desired signal below the effective quantization resolution. To model genuine nonlinear clipping/saturation without sufficient AGC protection, a nonlinear model would be required instead of a linear one. Such nonlinear clipping analysis is beyond the scope of this paper.}

\section{Coherent Jamming DoA Estimation}\label{3}
Due to the special structure of our proposed DARISAA, the receiver is able to suppress the high-power jamming at both the analog DARISAA end and the multi-channel digital processor end. However, the premise of jamming suppression and desired signal reception is to have a priori DoA information of both signals. Traditionally, DoA estimation should be performed in the digital signal processing (DSP) module of the receiver. 
\rev{However, in the suppressive jamming scenario considered in this paper, the jamming may be 90 dB stronger than the desired signal. Under AGC operation, such strong jamming dominates the ADC dynamic range and compresses the weak desired signal below the effective quantization resolution before the mixed signals reach the DSP. As a result, the low-power desired signal cannot be effectively detected or estimated in the DSP module.}
This leads to a paradox: to eliminate jamming and realize reliable desired signal reception we have to estimate their DoAs in the DSP module, \rev{but the jamming-dominated AGC and quantization process may render the weak desired signal unresolvable, so the desired-signal DoAs cannot be correctly estimated in the DSP module.}

To break this paradox, jamming must be effectively eliminated before it reaches the AGC and ADC so the DoA estimation in the DSP can be correctly performed. In our DARISAA-enabled receiver, this goal can be realized by exploiting the flexible adjustability of the responses of the DARISAA metasurface elements, i.e., the DARISAA is able to directly eliminate the jamming before it enters the RF chain and DSP.

The whole procedure consists of the following three steps:

1) When the mixed signals impinge on the DARISAA and then enter the RF chains, the AGCs and ADCs adjust to adapt to the dynamic range of the mixed signals, which is dominated by the jamming signal, and then the DoAs of jamming signal are estimated in the DSP module. Note that in this stage, due to the multipath effect, there are multiple DoAs of the same jamming signal, which is a DoA estimation problem of multiple coherent sources. We will propose a DoA estimation method by exploiting the dynamic agile adjustment capability of the DARISAA. 

2)  With the DoAs of the jamming signal, we then adjust the phase response of the DARISAA to eliminate jamming from these DoAs in the spatial domain. This  realizes the analog filtering in the antenna domain and greatly weakens the jamming power before it enters the AGCs and ADCs. 
\rev{The dynamic range of the mixed signals containing the residual jamming and the desired signal will be significantly reduced, thereby alleviating AGC compression and improving the effective quantization resolution for the desired signal. Consequently, the desired signal can be reliably quantized and passed to the digital processor.}
Then, the DoAs of the desired signal can be effectively estimated in the DSP module again similarly to that in the first step.  

3) With both the DoAs of jamming and desired signals, we jointly optimize the DARISAA response $\bf V$ and the digital anti-jamming beamformer $\bf w$ to maximize the SINR of the output desired signal at the receiver.

Overall, the proposed comprehensive scheme involves the following three steps: coherent jamming DoA estimation, DARISAA response optimization to eliminate jamming signal and coherent desired signal DoA estimation, and SINR maximization via joint optimization of DARISAA response and digital beamformer.

\subsection{Challenges of DoA  Estimation of Coherent Source}
In the first step, we estimate the DoAs of the jamming signal. Since
multipaths with different DoAs are coming from the same
jamming source, this is a DoA estimation of coherent signals. It
is well known that subspace-based DoA estimation algorithms
such as the multiple signal classification (MUSIC) algorithm \cite{15}
and estimation of signal parameters via rotational invariance
techniques (ESPRIT) \cite{16}, can be employed to estimate the DoA
with super-resolution, but they fail with coherent signals because the constructed covariance matrix is rank-deficient. The classical countermeasures such as the forward spatial smoothing (FSS) methods \cite{17} suffer from the drawback of aperture loss and deteriorating estimation accuracy. In this subsection, we utilize the dynamic agile adjustment property of our DARISAA and propose a new time-switched smoothing (TSS) method.

\rev{We construct the covariance matrix ${\mathbf{R}}_z$  of the received signals ${\bf{z}}$ in Eq. (\ref{eq3}) as}
\begin{equation}	\label{eq6} 		
	\begin{split}
		{\bf{R}}_z&=\frac{1}{L}\sum\limits_{l = 1}^L {{\mathbf{z}}\left( l \right){{\mathbf{z}}^H}\left( l \right)} \\
		&={(1 - \alpha )^2}P_J{{\bf{V}}^H}{{\bf{A}}_J}{{\bm{\beta }}_J}{\bm{\beta }}_J^H{\bf{A}}_J^H{\bf{V}} + {(1 - \alpha )^2}\sigma_n^2\bf{I}
		 \\	& \ \ \ +{(1 - \alpha )^2}P_s{{\bf{V}}^H}{\bf{A}}{\bm{\beta }}{{\bm{\beta }}^H}{{\bf{A}}^H}{\bf{V}}+{\bf{R}}_{n_q},
	\end{split}
\end{equation}
where $l$ denotes the snapshot index of the received data symbols, $L$ is the number of snapshots, and $P_s$ and $P_J$ represent the power of the desired signal and  the jamming, respectively. At this point, since the SIR is very small, the jamming dominates the received signal. The $K$ jamming signals are perfectly coherent multipath components originating from a single source, resulting in a source covariance matrix with a rank of one. Therefore, directly applying conventional subspace-based DoA estimation methods is not feasible.

For the FSS method \cite{18, 19}, the covariance matrix is divided into submatrices and then they are averaged to recover the rank of coherent signals. In contrast, the ESPRIT-like methods \cite{20, 21} decorrelate coherent signals by extracting specific correlation elements from the original sample covariance matrix to mathematically reconstruct a new full-rank equivalent covariance matrix. However, this mathematical reconstruction fundamentally reduces the dimension of the equivalent matrix to approximately half of the original array manifold, which successfully resolves the coherence issue but inevitably results in a reduction of the effective spatial aperture by half. 

\rev{Instead of sacrificing spatial aperture, temporally varying configurations have also been exploited for IRS-assisted angle estimation.  In \cite{WangLiu2023}, a passive IRS is used as an external reflecting surface, and multiple IRS reflection patterns are applied over different time samples together with transmitter-side repetition coding. This creates a temporal-domain virtual array response, whose rank condition enables MUSIC-based DoA estimation for the user-to-IRS link. In this paper, a similar time-domain multi-observation principle is tailored to the sub-connected hybrid analog-digital DARISAA receiver architecture under suppressive jamming. The proposed TSS method exploits the fast intra-symbol reconfiguration of DARISAA to observe the same received symbol under multiple receive antenna response patterns, and restores the covariance rank of coherent signals by averaging the resulting covariance matrices.  Unlike \cite{WangLiu2023}, TSS exploits the intra-symbol agility of DARISAA to obtain multiple receiver-side observations of the same symbol, thereby requiring no transmitter-side repetition coding, and is further integrated with pre-ADC analog jamming suppression.}

\subsection{Covariance Matrix Rank Restoration Using TSS}
Consider that there are $T_p$ agile switching states within a single symbol period, which means that the response pattern of the DARISAA $\bf V$ adjusts $T_p$ times to obtain the receive signals in one symbol duration. 
\rev{For TSS to be valid, the $T_p$ agile states within one symbol period are assumed to observe the same transmitted symbol under different DARISAA configurations. Moreover, because the covariance matrix is estimated from $L$ snapshots, the DoAs and path coefficients are assumed to be block-wise quasi-static over the covariance estimation interval.}
Let ${{{\mathbf{V}}^{\left( t_p \right)}}}$ represent the analog phase response matrix of the $t_p$-th agile state, where $t_p=1,2,\ldots,T_p$. Consequently, for each snapshot $l$, we obtain $T_p$ distinct observations. We assume that all DARISAs at one snapshot apply the same randomly generated response vector ${{\mathbf{v}}^{\left( t_p \right)}} = {\mathbf{v}}_{1}^{\left( t_p \right)} =  \cdots  = {\mathbf{v}}_{{N_R}}^{\left( t_p \right)} = 1/\sqrt {{N_E}} {\left[ {\mathbf{v}^{\left( t_p \right)}(1), \mathbf{v}^{\left( t_p \right)}(2),\ldots,\mathbf{v}^{\left( t_p \right)}(N_E)} \right]^T}$. This special design will be used to estimate the DoAs with our proposed method, as will be detailed in the following subsections.

Because the suppressive jamming power is very high, the powers of the desired signal and noise are negligible.
\rev{Moreover, for a fixed snapshot index $l$, only $\mathbf{V}^{(t_p)}$ changes with $t_p$, while the symbol and propagation parameters remain common to all $T_p$ observations.}
 Therefore, the received signal of one snapshot can be expressed as
\begin{equation}
	\label{deqn_ex19a}
	{\mathbf{z}}^{\left( t_p \right)}(l) =\left( {1 - \alpha } \right) {\left( {{{\mathbf{V}}^{\left( t_p \right)}}} \right)^H}{\mathbf{A}}_J{\bm{\beta }_J}s_J\left( l \right) + {\mathbf{o}}^{\left( t_p \right)}\left( l \right),
\end{equation}
where ${\bf{o}}^{\left( t_p \right)}\left( l \right) = \left( {1 - \alpha } \right){\left( {{{\mathbf{V}}^{\left( t_p \right)}}} \right)^H}\left( {{\bf{A}\bm{\beta }}s + {\bf{n}}} \right) + {{\bf{n}}_q}$ represents the equivalent noise.

Due to the ULA configuration, each ${\bf{a}}(\theta _J^n), n=1,\ldots, K$ has a Vandermonde form, so the received signals of the DARISAA in (\ref{deqn_ex19a}) could be further written as
\begin{equation}
		\label{deqn_ex8}
		{{\mathbf{z}}^{\left( t_p \right)}}\left( l \right) = {\mathbf{\tilde A}_J}{{\mathbf{U}}^{\left( t_p \right)}}{\bm{\beta }_J}s_J\left( l \right) + {\mathbf{o}}^{\left( t_p \right)}\left( l \right),
\end{equation}	
where ${\mathbf{\tilde A}}_J = (1-\alpha) \left[ {{{{\mathbf{\tilde a}}}\left( { {\theta}_J^1 } \right)},\ldots,{{{\mathbf{\tilde a}}}\left( { {\theta}_J^K } \right)}} \right]\in\mathbb{C}^{{N_R} \times K}$ is the equivalent array manifold matrix and 
\begin{equation}
		\label{deqn_ex9}
		{{{\mathbf{\tilde a}}}\left( { {\theta}_J^k } \right)} = {\left[ {1,{{\rm{e}}^{ - j\omega (\theta_J^k )}},\ldots,{{\rm{e}}^{ - j\left({N_R-1}\right)\omega (\theta_J^k )}}} \right]^T},
\end{equation}
with $\omega (\theta_J^k ) = 2\pi {N_E}\frac{d}{\lambda }\sin \theta_J^k$, and ${{\mathbf{U}}^{\left( t_p \right)}} $ is a diagonal matrix given by 
\begin{equation}
		\label{deqn_ex10}
		{{\mathbf{U}}^{\left( t_p \right)}} = \operatorname{diag}\left( {u_{}^{\left( t_p \right)}\left( {{\theta _J^1}} \right),\ldots,u_{}^{\left( t_p \right)}\left( {{\theta _J^K}} \right)} \right),
\end{equation}
with $	{{u^{\left( t_p \right)}}\left( {{\theta _J^k}} \right)}=\sum\limits_{{n_e} = 1}^{{N_E}} {\mathbf{v}^{( t_p )}(n_e){e^{ - j2\pi {\left({{n_e}-1}\right)}\frac{d}{\lambda }\sin {\theta _J^k}}}}$.

 Specifically, because the $K$ incident jamming signals are fully coherent, their source covariance matrix is severely rank-deficient. To restore the rank of the signal covariance matrix, we perform smoothing across the $T_p$ agile states. The smoothed covariance matrix $\mathbf{\bar{R}}$ is computed by averaging over both the $L$ snapshots and the $T_p$ agile states:
\begin{equation}
		\label{deqn_ex15}
		\begin{aligned}
			{\mathbf{\bar R}} &= \frac{1}{T_{p}L}\sum\limits_{t_p = 1}^{T_p}\sum_{l=1}^{L} {{{\mathbf{z}}^{\left( t_p \right)}(l)}{{\left( {{{\mathbf{z}}^{\left( t_p \right)}}(l)} \right)}^H}} \\
			&	 = \frac{P_J}{T_p}{\mathbf{\tilde A}_J}\sum\limits_{t_p = 1}^{T_p} {{{\mathbf{U}}^{\left( t_p \right)}}{\bm{\beta }_J}{{\bm{\beta }_J}^H}{{\left( {{{\mathbf{U}}^{\left( t_p \right)}}} \right)}^H}} {{{\mathbf{\tilde A}_J}}^H} + {\mathbf{\bar R}_o}\\
			&	= \frac{P_J}{T_p}{\mathbf{\tilde A}_J}{{{\mathbf{\bar R}}}_J}{{{\mathbf{\tilde A}_J}}^H} +  {\mathbf{\bar R}_o},
		\end{aligned}
\end{equation}
where
\begin{equation}
	\begin{aligned}
		{\mathbf{\bar R}_o} &= \frac{1}{T_{p}L}\sum\limits_{t_p = 1}^{T_p}\sum_{l=1}^{L} {{{\mathbf{o}}^{\left( t_p \right)}(l)}{{\left( {{{\mathbf{o}}^{\left( t_p \right)}}(l)} \right)}^H}},
		\end{aligned}
	\end{equation}
	and
\begin{equation}
		\label{deqn_ex16}
		\begin{aligned}
				{{{\mathbf{\bar R}}}_J} 
				= \sum\limits_{t_p = 1}^{T_p} {\left( {{{\mathbf{U}}^{\left( t_p \right)}}{\bm{\beta }}_J} \right){{\left( {{{\mathbf{U}}^{\left( t_p \right)}}{\bm{\beta }}_J} \right)}^H}} 
				= {\mathbf{C}}{{\mathbf{C}}^H},
			\end{aligned}
\end{equation}
and
\begin{equation}
		\label{deqn_ex17}
				{\mathbf{C}} = \left[ {{{\mathbf{U}}^{\left( 1 \right)}}{\bm{\beta }_J},\ldots,{{\mathbf{U}}^{\left( T_p \right)}}{\bm{\beta }}_J} \right]= {\mathbf{B}}	{{\mathbf{U}}},
\end{equation}
with ${\mathbf{B}} = {\rm{diag}}\left( {\beta _J^1,\beta _J^2,\ldots,\beta _J^K} \right)$,  ${{\mathbf{U}}} = {\left[ {{{\mathbf{u}}^{\left( 1 \right)}},{{\mathbf{u}}^{\left( 2 \right)}},\ldots,{{\mathbf{u}}^{\left( T_p \right)}}} \right]}$, and $\forall t_p$, ${{\mathbf{u}}^{\left( t_p \right)}} = {\left[ {u_{}^{\left( t_p \right)}\left( {{\theta _J^1}} \right),\ldots,u_{}^{\left( t_p \right)}\left( {{\theta _J^K}} \right)} \right]^T}$.

From the above analysis, ${\mathbf{B}}$ is a diagonal matrix whose rank is $K$. ${{\mathbf{u}}^{\left( t_p \right)}}$ is a column vector formed by the elements on the diagonal of the matrix ${{\mathbf{U}}^{\left( t_p \right)}}$. It follows that $\mathrm{rank}\left( {\mathbf{C}} \right)=\min \left\{ {\mathrm{rank}\left( {\mathbf{B}} \right),\mathrm{rank}\left( {\mathbf{U}} \right)} \right\} = \min \left\{ {K, T_p} \right\}$,  provided that the agile response vectors ${{\mathbf{u}}^{\left( t_p \right)}}$ are chosen to be linearly independent. Therefore, we can get 
$\mathrm{rank}\left( {{{{\mathbf{\bar R}}}_J}} \right) = \mathrm{rank}\left( {\mathbf{C}} \right)= \min \left\{ {K, T_p} \right\}$. It should be noted that when $T_p \ge K$, the rank of ${{{\mathbf{\bar R}}}_J}$ is restored to $K$. When $T_p=1$, the proposed TSS method is ineffective, and the rank of ${{{\mathbf{\bar R}}}_J}$ drops to 1. This process avoids dividing the covariance matrix into overlapping subarrays and thus avoids array aperture loss, which maintains the estimation accuracy.

{\emph {Remark 2:}}
The identical response constraint across DARISAs (i.e., $\mathbf{v}^{(t_p)} = \mathbf{v}_1^{(t_p)} = \dots = \mathbf{v}_{N_R}^{(t_p)}$) is exclusively enforced during the DoA estimation stage to enable subspace-based algorithms. In the subsequent ambiguity resolution and hybrid beamforming stages, this constraint is completely relaxed, thereby fully restoring the system's spatial degrees of freedom.

\subsection{Virtual DoA Estimation Using Root-MUSIC}			
The jamming DoAs can be estimated from the smoothed covariance matrix ${{\mathbf{\bar R}}}$ using subspace-based algorithms. The eigenvalue decomposition of ${{\mathbf{\bar R}}}$ is given by
\begin{equation}\label{deqn_ex18}
		{\mathbf{\bar R}} = \left[ {\begin{array}{*{20}{c}}
					{{{\mathbf{U}}_J}}&{{{\mathbf{U}}_N}}
			\end{array}} \right]{\mathbf{\Lambda}} {\left[ {\begin{array}{*{20}{c}}
						{{{\mathbf{U}}_J}}&{{{\mathbf{U}}_N}}
		\end{array}} \right]^H},
\end{equation}
where ${{{\mathbf{U}}_J}}$ is the signal subspace which is composed of the eigenvectors corresponding to the $K$ largest eigenvalues, and ${{{\mathbf{U}}_N}}$ consists of the eigenvectors corresponding to the smallest ${N_R}-K$ eigenvalues, which denote the noise subspace. Since the virtual array is a ULA, we can employ the Root-MUSIC algorithm \cite{22}, which offers higher accuracy by finding the roots of a polynomial rather than performing a spectral search. A matrix $\mathbf{Q}$ is first constructed from the noise subspace eigenvectors ${{\mathbf{U}}_N}$, which is expressed as ${\mathbf{Q}} = {{\mathbf{U}}_N}{\mathbf{U}}_N^H$.
The coefficients of a polynomial are then constructed by summing the elements along the diagonals of $\mathbf{Q}$. The $K$ roots of this polynomial that lie closest to the unit circle, denoted $\{r_J^k\}_{k=1}^K$, correspond to the jamming signals. The desired virtual phases, $\tilde\omega\left(\hat{\theta}_J^k\right)$, are estimated from the angles of these roots by
\begin{equation}
	\tilde\omega\left( \hat{\theta}_J^k\right) =\mathrm{angle}( r_J^k),\quad k=1,\ldots,K,
\end{equation}
where the $\mathrm{angle}(\cdot)$ operator extracts the principal phase angle of a complex number within the range $(-\pi, \pi]$.  Note that according to $\omega\left( \hat{\theta}_J^k\right)$  in ${\mathbf{\tilde a}}\left( {\hat{\theta}_J^k } \right)$ in  (\ref{deqn_ex9}), we should actually obtain the optimal $K$ solutions as
\begin{equation}
	\Omega  = \left\{ {{\omega\left( \hat{\theta}_J^k \right) }\mid k = 1,2,\ldots,K} \right\},
\end{equation}
and the actual spatial phase range of ${\omega\left( \hat{\theta}_J^k \right) }$ is $\left(-{\pi}{N_E},{\pi}{N_E} \right]$ (assuming $d=\lambda/2$).
 
However, the Root-MUSIC algorithm  returns only the principal value of the phase between $\left(-{\pi},{\pi} \right]$, which means that there is spatial phase aliasing. This is because the spacing between two adjacent RF chains is $N_E d = N_E \lambda/2$, due to the special structure of DARISAA. According to spatial sampling theory, this specific undersampling ratio introduces exactly $N_E$ periodic ambiguities within the visible angular region $[-\pi/2, \pi/2]$. Consequently, all $N_E$ unwrapped phases should be considered, and thus the associated set $\hat{\Theta}_{J}^{k}$ with 	$N_E$ candidate angles is expressed as
		\begin{equation}
			\label{deqn_ex22}
			\hat{\Theta}_{J}^{k} = \left\{ \sin^{-1}\left(\frac{\tilde\omega(\hat{\theta}_{J}^{k})+2\pi i}{2\pi N_{E}d/\lambda}\right) \;\middle|\; i \in \{I_{\min}^k, \dots, I_{\max}^k\} \right\},
		\end{equation}
where ${I_{\min}^k}=\left\lceil -{N_E}d/{\lambda }\;-{\tilde\omega \left( {\hat{\theta}_J^k} \right)}/{2\pi }\; \right\rceil$, ${I_{\max}^k}=\left\lfloor {N_E}d/{\lambda }\;-{\tilde\omega\left( {\hat{\theta}_J^k} \right)}/{2\pi }\; \right\rfloor$. The notations $\left\lceil \cdot  \right\rceil $ and $\left\lfloor \cdot  \right\rfloor $ represent the ceiling and floor functions, respectively. Consequently, the complete set of feasible DoAs is
		\begin{equation}
			\label{deqn_ex23}
			\hat{\Theta}_{J} = \bigcup_{k=1}^{K} \hat{\Theta}_{J}^{k}.
		\end{equation}
		
\subsection{Ambiguity Resolution for Jamming DoAs}
To resolve the ambiguity associated with each of the $K$ estimated virtual DoAs, we perform a path-wise search using a low-complexity spatial matched filter (SMF). This procedure is executed independently for each of the $K$ signal paths.

For a specific path $k$, we have its DoA candidate set $\hat{\Theta}_{J}^{k}$. For each candidate angle $\hat{\theta}_{J}^{k,i}$ in this specific set, we design the analog weighting matrix as ${\mathbf{V}}_{{{\hat \theta }_J^{k,i}}}$, whose phase shifting is
\begin{equation}
	\label{deqn_ex24}
	{\alpha _{{n_r},{n_e}}} = 2\pi \left[(n_r - 1)N_E + (n_e - 1) \right]  \frac{d}{\lambda }\sin {{\hat \theta }_J^{k,i}},
\end{equation}
which realizes an analog phase alignment at ${{\hat \theta }_J^{k,i}}$.
Correspondingly, the phase shift of the digital beamformer ${\mathbf{w}}_{{{\hat \theta }_J^{k,i}}}$ is
\begin{equation}
	{\mathbf{w}}_{{{\hat \theta }_J^{k,i}}} = \frac{1}{{\sqrt {{N_R}} }}\mathbf{1}_{N_R}. \label{deqn_ex25}
\end{equation}
With the analog and digital beamformer in (\ref{deqn_ex24}) and (\ref{deqn_ex25}), the received signal is
\begin{equation}
	\begin{aligned}
		y_{{{\hat \theta }_J^{k,i}}}\left( l \right) ={}& {\mathbf{w}}_{{{\hat \theta }_J^{k,i}}}^H \Big( \left( {1 - \alpha } \right){{\mathbf{V}}_{{{\hat \theta }_J^{k,i}}}^H} ({\bf{A}\bm{\beta }}s\left( l \right) + {{\bf{A}}_J}{{\bm{\beta }}_J}{s_J}\left( l \right) \\
		& + {\bf{n}}\left( l \right)) + {{\bf{n}}_q}\left( l \right) \Big).
	\end{aligned}
\end{equation}
The output power is then calculated by averaging the power of the received signal $y_{\hat{\theta}_{J}^{k,i}}(l)$ over $L$ snapshots, expressed as
\begin{equation}
	\label{deqn_ex31}
	P_{{{\hat \theta }_J^{k,i}}}^{} = \frac{1}{L}\sum\limits_{l = 1}^L \left|y_{{{\hat \theta }_J^{k,i}}}\left( l \right) \right|^2.
\end{equation}
The ambiguity for the $k$-th path is then resolved by identifying the candidate angle within its set $\hat{\Theta}_{J}^{k}$ that produces the maximum output power, and the desired DoA is
\begin{equation}
	\hat{\theta}_{J}^{k} = \underset{\theta \in \hat{\Theta}_{J}^{k}}{\arg\max} \ P_{\theta}.
\end{equation}
This process is repeated for all $K$ paths, ultimately yielding the unambiguous DoAs $\{\hat{\theta}_{J}^{1}, \hat{\theta}_{J}^{2}, \dots, \hat{\theta}_{J}^{K}\}$.

\section{Jamming Suppression and Coherent Desired Signal DoA Estimation}\label{4}
With the estimated DoAs of the high-power jamming $\{\hat{\theta}_J^k\}_{k=1}^K$, in this section we go to the second step to eliminate the jamming signals from these  $K$ DoAs by adjusting the phase responses of the DARISA elements. This weakens the jamming power before it enters the AGCs and ADCs, and thus the DoAs of the desired signal could be estimated.

\subsection{Jamming Suppression Problem Formulation} 
The objective in this step is to optimize the phase responses of the DARISA elements with the unit-amplitude constraint, thereby minimizing the jamming signals entering the RF chains and preventing ADC saturation. Physically, the jamming power entering the $n_r$-th RF chain is determined by the response vector ${\bf{v}}_{n_r}$ of the corresponding $n_r$-th DARISA and the jamming steering vector. We have to emphasize that we still need to estimate the DoAs $\theta_1,\ldots,\theta_P$ of the coherent multipath of the desired signal with the similar TSS method, it requires that the response vector of each DARISA be equal, i.e.,  ${\mathbf{v}} = {\mathbf{v}}_{1} =  \cdots  = {\mathbf{v}}_{N_R}$. 

Due to the ULA configuration, each ${\bf{a}}(\hat\theta _J^k), k=1,\ldots,K$ has a Vandermonde form. Therefore, defining the first $N_E$ elements of ${\bf{a}}(\hat\theta _J^k)$ as a subvector ${\bf{a}}_1( {{\hat\theta _J^k}} ) = {\left( {1,{e^{ - j2\pi d/\lambda \sin {\hat\theta _J^k}}},\ldots,{e^{ - j\left( {N_E - 1} \right)2\pi d/\lambda \sin {\hat\theta _J^k}}}} \right)^T} \in \mathbb{C}^{N_E\times 1}$, it is readily apparent that ${\bf{a}}_{n_r}\left( {{\hat\theta _J^k}} \right) = e^{-j(n_r-1)N_E2\pi d/\lambda \sin {\hat\theta _J^k}} {\bf{a}}_1\left( {{\hat\theta _J^k}} \right), n_r=1,\ldots, N_R$. With all the above considerations, the jamming suppression problem can be formulated as
\begin{align}
\label{eq11}
\mathop {\min } \limits_{{\bf{v}}} \quad & g_1({\bf{v}}) =\sum\limits_{k = 1}^K \left| {{\bf{v}}^H}{\bf{a}}_1(\hat\theta _J^k)\right|^2, \\
\mbox{s.t.}\quad
&\left| {{\bf{v}}(n_e)} \right| = 1, n_e = 1,2,\ldots, {N_E}, \nonumber
\end{align}
where only a single DARISA vector ${\bf{v}}$ needs to be optimized because the proposed TSS method requires that the response vector of each DARISA be identical.

\subsection{Proposed MO Algorithm} \label{MO Algorithm}
The function $g_1({\bf{v}})$ with variable ${\bf{v}}$  is nonconvex, which is difficult to solve. In fact, define the set in the constraints as ${{\cal S}} = \left\{ {{\bf{v}}:{\rm{ }}\left| {{{\bf{v}}}(n_e)} \right| = 1,\forall n_e} \right\}$. We can see that the constraints for the variable ${\bf{v}}$ is in the manifold space  ${\cal S}$ and we can solve this complex problem based on the theory of manifold optimization (MO).
In this paper, we propose the MO algorithm based on the Riemannian conjugate gradient algorithm to optimize ${\bf{v}}$ .

The main idea of the MO algorithm is to search for a sequence of $ {{\bf{v}}_n} ,n = 0,1, \cdots,  $ in the  manifold ${\cal S}$, such that the objective function ${g_1}$ is optimized iteratively. When convergence is reached, the MO algorithm returns a solution that is the point at which the Riemannian gradient is zero. The solution is a suboptimal solution to problem (\ref{eq11}). The main advantage of this algorithm is that it can directly deal with non-convex (without any approximation) complex problems by means of the  theory of manifold optimization \cite{23} and it requires lower computational complexity compared to traditional methods.

We first give the tangent space of the manifold ${\cal S}$ at the point $ {{\bf{v}}} $.  For any point ${\bf{v}} \in {{\cal S}}$, the tangent space passes tangentially through ${\bf{v}}$. The corresponding tangent space  can be denoted as ${T_{\bf{v}}}{\cal S}$
where ${T_{\bf{v}}}{{\cal S}} = \left\{ {{{\bf{q}}} \in {\mathbb{C}^{{N_E} \times 1}}\left| {{\mathop{\rm Re}\nolimits} \left\{ {{{\bf{q}}} \odot {{\bf{v}}^*}} \right\} = {\bf{0}}} \right.} \right\}$, and  ${{\bf{q}}}$ represents a tangent vector at point $\bf{v}$. The tangent space $T_{\mathbf{v}}\mathcal{S}$ constitutes the set of all tangent vectors to the manifold at the point $\mathbf{v}$. It can be derived by differentiating the modulus constraint $|\mathbf{v}(n_e)|^2 = \mathbf{v}(n_e) \mathbf{v}(n_e)^* = 1$. Specifically, for any smooth curve on the manifold passing through point $\mathbf{v}$ with tangent vector $\mathbf{q}$, the differentiation of this constraint yields $\mathbf{q}(n_e) \mathbf{v}(n_e)^* + \mathbf{v}(n_e) \mathbf{q}(n_e)^* = 2\mathrm{Re}\{\mathbf{q}(n_e) \mathbf{v}(n_e)^*\} = 0$. Consequently, the tangent space is defined as the set of vectors $\mathbf{q}$ satisfying $\mathrm{Re}\{\mathbf{q} \odot \mathbf{v}^*\} = \mathbf{0}$.

Under the concept of tangent space, the next key part of the MO algorithm is to calculate the Riemannian gradient, which is the tangent direction in which the objective function decreases the fastest among all tangent vectors in the tangent space. Since ${{\cal S}} \in {\mathbb{C}^{{N_E} \times 1}}$ is a submanifold of the Euclidean space, according to \cite{23}, the Riemannian  gradient can be computed as the orthogonal projection of the conventional Euclidean gradient onto the Riemannian tangent space. Specifically,  it can be written as
\begin{equation}
\label{eq30}
gra{d_{{{\bf{v}}_n}}}{g_1} = Gra{d_{{{\bf{v}}_n}}}{g_1} - {\mathop{\rm Re}\nolimits} \left\{ {Gra{d_{{{\bf{v}}_n}}}{g_1} \odot {\bf{v}}_n^*} \right\} \odot {{\bf{v}}_n},
\end{equation}
where ${{\bf{v}}_n}$ represents the  results obtained in the $n$-th iteration, $gra{d_{{{\bf{v}}_n}}}{g_1}$ is the Riemannian gradient of ${{\bf{v}}_n}$, and $Gra{d_{{{\bf{v}}_n}}}{g_1}$  is the Euclidean gradient of ${{\bf{v}}_n}$. The Euclidean gradient of the objective function  $g_1$ with respect to  ${{\bf{v}}_n}$ can be calculated as
\begin{equation}
\label{eq35}
Gra{d_{{{\bf{v}}_n}}}{g_1} = \sum\limits_{k=1}^{K}2{{\bf{a}}_1}(\hat\theta _{J}^{k}){{\bf{a}}_1}{{(\hat\theta _{J}^{k})}^{H}}{\bf{v}}_n.
\end{equation}

With the Riemannian gradient, the next step is to find the search direction at the point $ {{{\bf{v}}_n}} $. The search direction  ${{\bf{\Pi }}_{{{\bf{v}}_n}}}$ at $ {{{\bf{v}}_n}} $ should be updated at the iteration $n$  as
\begin{equation}
	\begin{split}
		\label{eq43}
		{{\bf{\Pi }}_{{{\bf{v}}_n}}} = - gra{d_{{{\bf{v}}_n}}}{g_1} + \mu _n^{{\bf{v}}}{T_{{{\bf{v}}_{n - 1}} \to {{\bf{v}}_n}}}\left( {{{\bf{\Pi }}_{{{\bf{v}}_{n - 1}}}}} \right),
	\end{split}
\end{equation}
where ${T_{{{\bf{v}}_{n - 1}} \to {{\bf{v}}_n}}}\left( {{{\bf{\Pi }}_{{{\bf{v}}_{n - 1}}}}} \right)$ represents  $ {{{\bf{\Pi }}_{{{\bf{v}}_{n - 1}}}}} $ on the manifold ${T_{{{{\bf{v}}_{n - 1}}} }}{\cal S}$ to the manifold ${T_{ {{{\bf{v}}_n}} }}{\cal S}$, which is defined as
\begin{equation}
	\begin{split}
		{T_{{{\bf{v}}_{n - 1}} \to {{\bf{v}}_n}}}\left( {{{\bf{\Pi }}_{{{\bf{v}}_{n - 1}}}}} \right) = {{\bf{\Pi }}_{{{\bf{v}}_{n - 1}}}} - {\mathop{\rm Re}\nolimits} \left\{ {{{\bf{\Pi }}_{{{\bf{v}}_{n - 1}}}} \odot {\bf{v}}_n^*} \right\} \odot {\bf{v}}_n,
		\label{eq45}
	\end{split}
\end{equation}
where $\mu _n^{{\bf{v}}}$ is the Polak-Ribière parameter defined  in \cite{23} as
\begin{equation}
\mu _n^{{\bf{v}}} = \frac{{{{\left\langle {gra{d_{{{\bf{v}}_n}}}{g_1},gra{d_{{{\bf{v}}_n}}}{g_1} - {T_{{{\bf{v}}_{n - 1}} \to {{\bf{v}}_n}}}\left( {gra{d_{{{\bf{v}}_{n - 1}}}}{g_1}} \right)} \right\rangle }}}}{{{{\left\langle {gra{d_{{{\bf{v}}_{n - 1}}}}{g_1},gra{d_{{{\bf{v}}_{n - 1}}}}{g_1}} \right\rangle }}}}.
\label{eq46}
\end{equation}

With the search direction  ${{\bf{\Pi }}_{{{\bf{v}}_n}}}$, the iteration update is now to calculate  ${{\bf{v}}_{n + 1}}$ as
\begin{align}
{{\bf{v}}_{n + 1}} ={\cal R}\left(  {{\bf{v}}_n} + {\lambda_n}{{\bf{\Pi }}_{{{\bf{v}}_n}}}\right), 	\label{eq48}
\end{align}
where ${\lambda_n}$ is the step size of the $n$-th iteration, which could be calculated using the back-tracking line search algorithm \cite{23}. Note that the elements in  ${{\bf{v}}_{n + 1}}$ may not satisfy the unit-modulus  constraint, and thus, a retraction operation is required during each iteration of the back-tracking line search, defined as
${\cal R}({\bf{x}}(j)) = \frac{{{\bf{x}}(j)}}{{\left| {{\bf{x}}(j)} \right|}}, \forall j.$
This ensures that each non-zero element has a unit modulus.

\begin{algorithm}[t]
	\caption{MO Algorithm for jamming suppression with DARISAA configuration}\label{alg:MO}
	\begin{algorithmic}[t]
		\REQUIRE: $\varepsilon  > 0$, $c,\tau  \in \left( {0,1} \right)$ and initial point ${{\bf{v}}_0}$
		\STATE Set $n=0$, $\mu _0^{{\bf{v}}}= 0$,  ${{\bf{\Pi }}_{{{\bf{v}}_0}}} =  - gra{d_{{{\bf{v}}_0}}}{g_1}$
		\REPEAT
		\STATE Calculate search step size ${\lambda _n}$ according to backtracking line search algorithm, and find the next point  ${{\bf{v}}_{n + 1}}$  by  (\ref{eq48}).  
		\STATE Find the Riemannian gradient of the new point  ${gra{d_{{{\bf{v}}_{n + 1}}}}{g_1}}$ according to  Eq. (\ref{eq30}).
		\STATE Calculate the vector transfer factor  ${T_{{{\bf{v}}_n} \to {{\bf{v}}_{n + 1}}}}\left( {{{\bf{\Pi }}_{{{\bf{v}}_n}}}} \right)$ through Eq. (\ref{eq45}).
		\STATE Calculate Polak-Ribière parameter  $\mu _{n + 1}^{{\bf{v}}}$ according to Eq. (\ref{eq46}).
		\STATE Calculate the search direction  ${{\bf{\Pi }}_{{{\bf{v}}_{n + 1}}}}$ by Eq. (\ref{eq43}).
		\STATE $n=n+1$.
		\UNTIL{$ {\left\| {gra{d_{{{\bf{v}}_n}}}{g_1}} \right\|} < \varepsilon $}
	\end{algorithmic}
\end{algorithm}

We summarize the proposed MO algorithm as Algorithm \ref{alg:MO}. In this algorithm, $c$ is the percentage decrease in the acceptable value of ${g_1}$ in the backtracking line search, $\tau $ is the parameter used to regulate the change in the step size $\lambda_n $, and $\varepsilon $ is used to control the accuracy of the Riemannian gradient at convergence.
When the algorithm starts with $n = 0$, we first set the initial search direction  ${{\bf{\Pi }}_{{{\bf{v}}_0}}} =  - gra{d_{{{\bf{v}}_0}}}{g_1}$,  and then compute the step size of this iteration by backtracking line search. Once the Riemannian gradient reaches the target accuracy $\varepsilon $, the optimization algorithm stops, otherwise the above steps are repeated until convergence.

Up to now, we obtain a solution of problem (\ref{eq11}). Note that the solution obtained by the MO algorithm is not the global optimum and the random initialization leads to different suboptimal solutions, each of which can greatly suppress the high-power jamming. We have successfully suppressed the primary jamming power at the analog front-end. The desired signal can now be quantized with sufficient effective resolution, enabling the estimation of its DoAs.

\subsection{Coherent Desired Signal DoA Estimation and Ambiguity Resolution}
Similar to the DoA estimation for jamming signals, the TSS method to estimate DoAs of the desired signal requires dynamic agile adjustment of the DARISAA response ${\bf V}^{(t_p)}$ to obtain $T_p$ observations during one symbol duration and to construct the covariance matrix,
thereby addressing the rank deficiency problem.
However, what is different is that, in this step all these ${\bf V}^{(t_p)}$ should satisfy the condition that they are able to eliminate the suppressive jamming so that the weak desired signal could enter the RF chains and ADCs. Obviously, this could be realized by using those suboptimal solutions of problem (\ref{eq11}), i.e., each ${\bf v}^{(t_p)}$ equals one solution of Algorithm \ref{alg:MO} with one random initialization. Then, ${\bf V}^{(t_p)}$ can be obtained by using formula (\ref{eq2}). Similar to the jamming signal DoA estimation, the TSS method is employed to obtain a set of ambiguous virtual angles $\hat \Theta$ corresponding to the desired signal paths.

However, a significant challenge arises in the disambiguation step. The simple SMF approach, which is effective for identifying the high-power jammer, is no longer viable. The fundamental limitation lies in the fact that an SMF configuration is designed only to maximize the gain towards a candidate angle, without any mandate to suppress jamming from jamming directions. Consequently, with the residual jamming power still being substantially stronger than the desired signal, the SMF output is dominated by jamming leakage through its sidelobes. This leakage masks the true signal peak, leading to a disambiguation failure.

Furthermore, a more fundamental restriction prevents this task from being performed in the digital domain. From the perspective of the digital processor, which only accesses the $N_R$ outputs of the RF chains, the system appears as a sparse virtual array with a large inter-antenna spacing of $N_Ed$. This significant undersampling is the source of the DoA ambiguity. Therefore, the ambiguity is inherent to the digital data itself and cannot be resolved using only these $N_R$ channels.

To address the limitations mentioned above, we propose a novel strategy that resolves the ambiguity in the analog domain by dynamically reconfiguring the antenna aperture. We first introduce a network of controllable RF switches that can, for the duration of the disambiguation process, connect all $N$ metasurface elements to a single RF chain. This operation synthesizes a single, large-aperture, $1/2\lambda$ space sampled ULA, providing the high spatial resolution necessary to unambiguously identify the true DoAs. Then, a more robust disambiguation method based on jamming-aware constrained beamforming is proposed. Instead of merely matching to a candidate angle as in (\ref{deqn_ex25}), this spatial filtering technique actively designs a beam pattern for each candidate that simultaneously preserves gain toward the candidate direction while enforcing deep nulls in the known jamming directions. Specifically, for the $i$-th candidate angle $\hat \theta_p^i$ in the ambiguous set $\hat \Theta$, we design an analog beamforming vector $\mathbf{\bar{v}} \in \mathbb{C}^{N \times 1}$ by solving the following optimization problem
	\begin{align}
		\mathop {\min } \limits_{{\bf{\bar{v}}}} \quad & g_2({\bf{\bar{v}}}) =\sum\limits_{k = 1}^K \left| {{\bf{\bar{v}}}^H}{\bf{a}}(\hat\theta _J^k)\right|^2, 	\label{ALM} \\
		\mbox{s.t.}\quad
		& \left| \mathbf{\bar{v}}^H \mathbf{a}(\hat \theta_p^i) \right|^2 \ge \delta, \nonumber \\
		&\left| {{\bf{\bar{v}}}(n)} \right| = 1, n = 1,2,\ldots, N. \nonumber
	\end{align}
The objective function minimizes the received power from the known jamming directions, effectively steering deep nulls. The first constraint preserves the gain towards the candidate signal direction above a threshold $\delta$. 

To solve the non-convex optimization problem presented in (\ref{ALM}) for the disambiguation vector ${\bf{\bar{v}}}$, we employ a hybrid approach that synergizes the augmented Lagrangian method (ALM) \cite{ref24, ref25} with MO. The ALM is first utilized to incorporate the signal gain inequality constraint into the objective function. We define the constraint violation function as $h(\mathbf{\bar{v}}) = \delta - |\mathbf{\bar{v}}^H \mathbf{a}(\hat \theta_p^i)|^2 \le 0$. The original problem is then transformed into a sequence of optimization problems by constructing the augmented Lagrangian function
\begin{equation}
	\begin{aligned}
		L_c(\mathbf{\bar{v}}, \lambda) = g_2(\mathbf{\bar{v}}) + \frac{1}{2c} \left( \left( 	\max\{0, \lambda + c \cdot h(\mathbf{\bar{v}})\} \right)^2 - \lambda^2 \right),
	\end{aligned}
\end{equation}
where $g_2(\mathbf{\bar{v}}) = \sum_{k=1}^{K} |\mathbf{\bar{v}}^H \mathbf{a}(\hat{\theta}_J^k)|^2$ is the original objective function defined in (\ref{ALM}), while $\lambda$ and $c$ are the Lagrangian multiplier and penalty factor, respectively. The ALM proceeds iteratively, where each main iteration involves solving the following subproblem subject to the constant modulus constraint
\begin{equation}
	\begin{aligned}
		\mathbf{\bar{v}}_{(t+1)} = \arg\min_{\mathbf{\bar{v}}} L_{c_t}(\mathbf{\bar{v}}, \lambda_t) \quad 
		\mbox{s.t.}\quad |\mathbf{\bar{v}}(n)|=1, \forall n.
	\end{aligned}
\end{equation}
This subproblem is addressed using the MO framework detailed in Section \ref{MO Algorithm}. The core of this process is the computation of the Riemannian gradient of the new objective function $L_c$. We begin by deriving its Euclidean gradient $Grad_{\mathbf{\bar{v}}} L_c$, presented in (\ref{ALM gradient}).
\begin{figure*}[t!]
	\begin{equation}\label{ALM gradient}
		Grad_{\mathbf{\bar{v}}} L_c = \sum\limits_{k=1}^{K}2\mathbf{a}(\hat{\theta}_{J}^{k})\mathbf{a}(\hat{\theta}_{J}^{k})^{H}\mathbf{\bar{v}} - \max\{0, \lambda_t + c_t \cdot h(\mathbf{\bar{v}})\} \cdot 2\mathbf{a}(\hat \theta_p^i)\mathbf{a}(\hat \theta_p^i)^H \mathbf{\bar{v}}.
	\end{equation}
\end{figure*}
Following the procedure established in (\ref{eq30}), the Riemannian gradient is then obtained by orthogonally projecting the Euclidean gradient onto the tangent space of the complex circle manifold at point $\mathbf{\bar{v}}$, namely
\begin{equation}
	\begin{aligned}
		grad_{\mathbf{\bar{v}}} L_c = Grad_{\mathbf{\bar{v}}} L_c - \operatorname{Re} \{Grad_{\mathbf{\bar{v}}} L_c \odot \mathbf{\bar{v}}^{*}\} \odot \mathbf{\bar{v}}.
	\end{aligned}
\end{equation}

With the Riemannian gradient defined, the beamforming vector $\mathbf{\bar{v}}$ for each candidate angle is found iteratively using  Algorithm \ref{alg:MO}. By applying each resulting spatial filter to the received signal, the corresponding output power is calculated. The candidate angle yielding the maximum power is then identified as the true DoA, thus resolving the ambiguity for one signal path. By repeating this procedure, the complete set of coherent multipath DoAs for the desired signal, $\{\hat{\theta}_1, \hat{\theta}_2, \ldots, \hat{\theta}_P\}$, is estimated.

\section{Anti-jamming by hybrid DARISAA and Digital Beamforming}\label{5}
With both the estimated DoAs of jamming and desired signals, we now 
maximize the output SINR at the digital end of the receiver by jointly optimizing the response of the DARISAA $\bf V$ and the beamformer of the digital end  ${\bf{w}}$.

The desired signal in (\ref{eq5}) is $ \left( {1 - \alpha } \right){{\bf{w}}^H}{{\bf{V}}^H}{\bf A\bm{\beta }}s$ and its power is $\tilde P_s = P_s{(1 - \alpha )^2}\left( {{{\bf{w}}^H}{{\bf{V}}^H}{\bf A\bm{\beta }}{{\bm{\beta }}^H}{{\bf{A}}^H}{\bf{Vw}}} \right)$. Similarly, the residual jamming and noise  can be denoted as ${{\bf{w}}^H}\left(\left( {1 - \alpha } \right) {{{\bf{V}}^H}\left( {{{\bf{A}}_J}{{\bm{\beta }}_J}{{\bf{s}}_J}{\bf{ + n}}} \right) + {{\bf{n}}_q}} \right)$ and its power is ${\tilde P_J} = {(1 - \alpha )^2}\left( {P_J{{\bf{w}}^H}{{\bf{V}}^H}{{\bf{A}}_J}{{\bm{\beta }}_J}{\bm{\beta }}_J^H{\bf{A}}_J^H{\bf{Vw}} + {{\bf{w}}^H}{{\bf{V}}^H}{{\bf{R}}_n}{\bf{Vw}}} \right) + {{\bf{w}}^H}{{\bf{R}}_{{n_q}}}{\bf{w}}$. Then, the objective now is to optimize ${\bf{w}}$ and ${\bf{V}}$ to maximize SINR, which can be expressed as
\begin{align}
	\mathop {\max }\limits_{{\bf{V}}{\rm{,  }}{{\bf{w}}}}\quad  &\mathrm{SINR} \left({\bf{V}}, {\bf{w}}\right) =\dfrac{\phi(\mathbf{V}, \mathbf{w})}{\psi(\mathbf{V}, \mathbf{w})} , 		\label{eq12}  \\
	\mbox{s.t.}\quad
	&{{\bf{V}} = {\rm{blkdiag}}({{\bf{v}}_1},{{\bf{v}}_2},\ldots,{{\bf{v}}_{{N_R}}})}, \nonumber \\
	&\left| {{{\bf{v}}_{n_r}}(n_e)} \right| = 1,n_r = 1,2,\ldots,{N_R},n_e = 1,2,\ldots,{N_E}, \nonumber\\
	&\|\mathbf{w}\|^2=1, \nonumber
\end{align}
where $\phi(\mathbf{V}, \mathbf{w})=(1-\alpha)^2\mathbf{w}^H\mathbf{V}^H\mathbf{T}_S\mathbf{V}\mathbf{w}$, $\psi(\mathbf{V}, \mathbf{w})=(1-\alpha)^2\mathbf{w}^H\mathbf{V}^H\mathbf{T}_{J}\mathbf{V}\mathbf{w}+\alpha(1-\alpha)\mathbf{w}^H\mathrm{diag}(\mathbf{V}^H\mathbf{R}_{\text{src}}\mathbf{V})\mathbf{w}$. Here, $\mathbf{T}_S=P_s{\bf{\hat A}}\bm{\beta }\bm{\beta }^H\mathbf{\hat{A}}^H$, $\mathbf{T}_{J}=P_J{\bf{\hat A}}_{J}\bm{\beta }_J\bm{\beta }_J^H{\bf{\hat A}}_{J}^H+\mathbf{R}_n$, $\mathbf{R}_\text{src}=\mathbf{T}_S+\mathbf{T}_{J}$, and $\mathbf{R}_n=\sigma_n^2\mathbf{I}$. Note that ${\bf{\hat A}}$ and ${\bf{\hat A}}_{J}$ represent the estimated array manifold matrices of the desired and jamming signals, respectively. In this step, we do not require  $ {\mathbf{v}}_{1} =  \cdots  = {\mathbf{v}}_{N_R}$.
We will use alternating optimization (AO) to optimize ${\bf{w}}$ and ${\bf{V}}$. To optimize the digital beamformer 
$\bf{w}$ while $\bf{V}$ is fixed, we solve a generalized Rayleigh quotient (GRQ) problem, and for ${\bf{V}}$ we will use the MO algorithm mentioned in  Algorithm \ref{alg:MO}, as detailed in the following subsections. \footnote{Although the product Riemannian manifold approach described in \cite{add-r3} enables joint optimization of manifold-constrained coupled variables, we instead employ the AO method. Specifically, the subproblem for $\mathbf{w}$ in problem (\ref{eq12}) constitutes a GRQ, allowing AO to efficiently obtain the exact global optimum via generalized eigenvalue decomposition.}

\subsection{Digital Beamformer Optimization}
We first consider optimizing  the digital beamformer ${\bf{w}}$ when  ${\bf{V}}$ is assumed to be fixed.  Thus, the problem (\ref{eq12})  can be written as 
	\begin{align}
	\mathop {\max }\limits_{{\bf{w}}}\quad  &f_1(\mathbf{w})=\rm{SINR} \left({\bf{w}}\right),   \\
	\mbox{s.t.}\quad
	&\left\|\mathbf{w}\right\|^{2}=1.  \nonumber
\end{align}
It is easy to see that the objective function is a GRQ, therefore, the globally optimal solution for ${{\bf{w}}}$ is the generalized eigenvector (GEV) corresponding to the maximal generalized eigenvalue \cite{24}. 
Although an explicit closed-form expression is generally unavailable, the GEV problem can be efficiently solved using standard numerical algorithms.

\subsection{Optimization of DARISAA Phases}
In this subsection, we optimize ${\bf{V}}$  with fixed ${\bf{w}}$. Similarly, every element in ${\bf{V}}$ has to satisfy the unit modulus constraint, so it can also be solved using the MO algorithm in Algorithm \ref{alg:MO}.
The optimization problem can be expressed as
	\begin{align}
		\label{eq16}
		\mathop {\max }\limits_{{\bf{V}}}\quad  &f_2(\mathbf{V})=\rm{SINR} \left({\bf{V}}\right),   \\
		\mbox{s.t.}\quad
		&{{\bf{V}} = {\rm{blkdiag}}({{\bf{v}}_1},{{\bf{v}}_2},\ldots,{{\bf{v}}_{{N_R}}})}, \nonumber\\
		&\left| {{{\bf{v}}_{n_r}}(n_e)} \right| = 1,n_r = 1,2,\ldots,{N_R},n_e = 1,2,\ldots,{N_E}.  \nonumber
	\end{align}
For the maximization problem in (\ref{eq16}), the MO solver is applied with a Riemannian ascent direction. The repeated algorithmic details are omitted for brevity, and only the Euclidean gradient with respect to $\mathbf V$ is given as
\begin{align}
	Gra{d_{{\bf{V}}_n}}{f_2} &= {\mathbf{C_V}} \odot {{\bf{Q}}_1}, 	\label{meq72}
\end{align}
where the matrix ${\mathbf{C_V}}$ shown in (\ref{eq17}) represents the complex gradient components derived from the objective function. 
\begin{figure*}[ht] 
	\begin{equation}
			\label{eq17}
		 {\mathbf{C_V}}=
		 \frac{2(1-\alpha)}{\psi(\mathbf{V})^2}
		 \left[
		 (1-\alpha)
		 \left(\psi(\mathbf{V})\mathbf{T}_S-\phi(\mathbf{V})\mathbf{T}_{J}\right)
		 \mathbf{V}\mathbf{w}\mathbf{w}^H
		 -\alpha \phi(\mathbf{V})\mathbf{R}_\text{src}\mathbf{V}\mathbf{D}_{w}
		 \right].
	\end{equation}
	
	\centering
\end{figure*}
Its full derivation is detailed in Equation (\ref{eq17}), where $\mathbf{D}_{w}=\mathrm{diag}\left(|w_1|^2,|w_2|^2,\ldots,|w_{N_R}|^2\right)$.
And $\mathbf{Q}_1=\mathrm{blkdiag}\left(\mathbf{1}_{N_E}^{(1)},\ldots,\mathbf{1}_{N_E}^{(N_R)}\right)$ is a block diagonal matrix with the same matrix structure as ${\bf{V}}$. Then, the Riemannian  gradient can be written as
\begin{equation}
		gra{d_{{{\bf{V}}}_n}}{f_2} = Gra{d_{{{\bf{V}}_n}}}{f_2} - {\mathop{\rm Re}\nolimits} \left\{ {	Gra{d_{{{\bf{V}}_n}}}{f_2} \odot {\bf{V}}_n^*} \right\} \odot {{\bf{V}}_n}.
\end{equation}
Note that the optimization variable in this subproblem is a matrix, so the expression for the  Polak-Ribière parameter $\mu _n$  needs to be changed and expressed as

\begin{equation}
	\begin{split}
		\mu _n^{{\bf{V}}} = \frac{{{{\left\langle {gra{d_{{{\bf{V}}_n}}}{f_2},gra{d_{{{\bf{V}}_n}}}{f_2} - {T_{{{\bf{V}}_{n - 1}} \to {{\bf{V}}_n}}}\left( {gra{d_{{{\bf{V}}_{n - 1}}}}{f_2}} \right)} \right\rangle }_R}}}{{{{\left\langle {gra{d_{{{\bf{V}}_{n - 1}}}}{f_2},gra{d_{{{\bf{V}}_{n - 1}}}}{f_2}} \right\rangle }_R}}},
		\label{eq18}
	\end{split}
\end{equation}
where ${\left\langle {.,.} \right\rangle _R}$ is the Riemannian metric, detailed as

\begin{equation}
	\label{eq19}
	{\left\langle {{\bf{X}},{\bf{Y}}} \right\rangle _R} = {\mathop{\rm Re}\nolimits} \left( {{\rm{Tr}}\left( {{{\bf{X}}^H}{\bf{Y}}} \right)} \right).
\end{equation}

\rev{The proposed DARISAA-based anti-jamming scheme is developed under an ideal hardware and array-manifold model. In practical implementations, however, hardware imperfections and model mismatch may affect its performance.
For example, fluctuations and finite precision in the varactor tuning voltages may lead to phase errors, which can reduce the achievable analog null depth and increase residual jamming leakage. In addition, mutual coupling among densely packed metamaterial elements may distort the ideal array manifold assumed in the TSS-based DoA estimation and beamforming design, thereby degrading covariance rank restoration, ambiguity resolution, and jamming suppression performance. Therefore, robust calibration and mismatch-aware optimization are important directions for practical DARISAA implementations.}

\section{Complexity Analysis}\label{6}
The computational complexity of the proposed anti-jamming procedure is analyzed by breaking it down into its three primary stages. To reflect the different optimization problems, we let $I_{\text{mo}_1}$, $I_{\text{mo}_2}$, and $I_{\text{mo}_3}$ denote the number of iterations required for the manifold optimization solver at different stages. Similarly, $I_{\text{alm}}$ and $I_{\text{alt}}$ denote the iterations for the ALM and the final AO loop, respectively.

\subsection{Jamming DoA Estimation}
The initial estimation of jamming DoAs involves three main steps. First, the construction of the temporally smoothed covariance matrix requires $T_p$ snapshots, with each snapshot involving matrix-vector multiplications, leading to a complexity of $\mathcal{O}(T_p L N_R^2)$. Second, the subsequent TSS method is dominated by the eigenvalue decomposition (EVD) of the $N_R \times N_R$ covariance matrix, which has a complexity of $\mathcal{O}(N_R^3)$. Finally, the ambiguity resolution for the $K$ jamming paths is performed using a spatial matched filter. Since each virtual phase yields $N_E$ candidate angles, this stage requires $\mathcal{O}(K N_E^2 L)$ operations. Thus, the total complexity for this stage is $\mathcal{O}(T_p L N_R^2 + N_R^3 + K N_E^2 L)$.

\subsection{Jamming Suppression and Desired Signal DoA Estimation}
The TSS process now incorporates an optimization step within each of its $T_p$ agile snapshots to generate jamming suppressing patterns. The optimization of each pattern using the manifold method has a complexity of $\mathcal{O}(I_{\text{mo}_1} K N_E^2)$, and thus the complexity of generating $T_p$ interference-suppressing patterns is $\mathcal{O}(T_p I_{\text{mo}_1} K N_E^2)$. This is followed by the standard covariance matrix update and EVD, contributing $\mathcal{O}(T_p L N_R^2 + N_R^3)$. The most demanding step is ambiguity resolution for the $P$ desired signal paths. For each of the $P \times N_E$ candidate angles, a constrained optimization problem is solved using an ALM-based manifold optimization approach. The dominant cost within this solver is the gradient calculation, which involves products with $N \times N$ matrices, resulting in a complexity of $\mathcal{O}(N^2)$ per inner iteration. Therefore, the complexity of this ambiguity resolution step is $\mathcal{O}(P N_E I_\text{alm} I_{\text{mo}_2} N^2)$.

	{\emph {Remark 3:}}
	 The computationally intensive manifold optimization is decoupled from the nanosecond-level hardware switching. Since the $T_p$ jamming suppression patterns are pre-calculated and merely applied sequentially during the desired signal DoA estimation, the optimization latency does not compromise real-time agility.

\subsection{Hybrid Analog-Digital Beamforming}
The final stage employs an AO algorithm to jointly find the analog beamformer $\mathbf{V}$ and the digital beamformer $\mathbf{w}$. For a fixed $\mathbf{w}$, the complexity of optimizing the block diagonal $\mathbf{V}$ using manifold optimization is $\mathcal{O}(I_{\text{mo}_3} N^2)$. For a fixed $\mathbf{V}$, optimizing $\mathbf{w}$ involves solving an $N_R \times N_R$ generalized eigenvalue problem, with a complexity of $\mathcal{O}(N_R^3)$. The total complexity for this final stage is therefore $\mathcal{O}(I_\text{alt} (I_{\text{mo}_3} N^2 + N_R^3))$.

In summary, the overall complexity is $\mathcal{O}(T_p L N_R^2 + N_R^3 + K N_E^2 L) + \mathcal{O}(P N_E I_\text{alm} I_{\text{mo}_2} N^2) + \mathcal{O}(I_\text{alt} (I_{\text{mo}_3} N^2 + N_R^3))$.

\section{Simulation Results}\label{7}
In this section, we present comprehensive simulation results to validate the performance of the proposed DARISAA anti-jamming system. We comprehensively evaluate the key stages of our scheme, including DoA estimation for both jamming and desired signals, analog-domain jamming suppression, and the final SINR improvement achieved through joint analog-digital beamforming.

\begin{figure}[!t]
	\centering
	\includegraphics[width=0.8\linewidth]{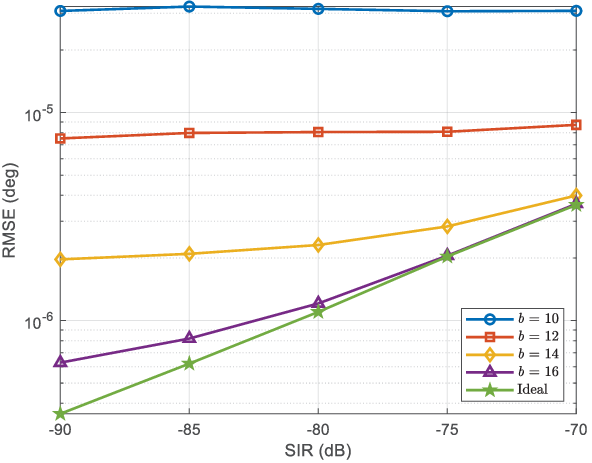}
	\caption{RMSE of jamming DoA estimation for different ADC resolutions.}
	\label{RMSE of jamming signal with different quantization bits}
\end{figure}

\subsection{Simulation Setup}
Unless otherwise specified, we consider a ULA receiver equipped with the following parameters: the receiver is equipped with $N_R=8$ DARISAs (i.e., 8 RF chains), each comprising $N_E=8$ reconfigurable elements, and each RF chain is followed by a 10-bit ADC. The 10-bit ADC imposes a critical input SIR threshold of approximately -56 dB for successful signal digitization, as shown in Fig. \ref{fig_adc}. Additionally, the carrier frequency is set to $f_c = 3$ GHz, the number of snapshots is $L = 128$, and the time-smoothing parameter for TSS is set to $T_p = 4$. The spacing between adjacent elements is set to $d=\lambda/2$. The simulation scenario involves two coherent desired signal paths arriving from $-20^{\circ}$ and $35^{\circ}$, and two coherent jamming paths from $-40^{\circ}$ and $55^{\circ}$. We assume, without loss of generality, that the signal paths have identical power distribution. Thus, each element of the channel coefficient vector ${\bm{\beta}}_J$ is $\beta_J^k=1/\sqrt{K}$, and similarly, $\beta_p=1/\sqrt{P}$. The input SNR and SIR are fixed at 20 dB and -90 dB, respectively. All results are averaged over 1000 Monte Carlo trials.

	{\emph {Remark 4:}}
	DARISAA's dynamic agility is highly feasible with current hardware. Specifically, the TSS performs $T_p$ state switches within a single symbol duration $T_s$. For a 10 MHz bandwidth (i.e., $T_s = \SI{100}{\nano\second}$) and $T_p = 4$, each agile state lasts $\SI{25}{\nano\second}$. Considering a varactor diode settling time of $1\sim10 \ \si{\nano\second}$, a sufficient window of over $\SI{15}{\nano\second}$ remains for ADC sampling. 
	Furthermore, for $L = 128$ snapshots, the covariance estimation interval is $L T_s=\SI{12.8}{\micro\second}$, which is much shorter than typical channel coherence times at the considered carrier frequency and therefore supports the block-wise quasi-static assumption used by TSS.

\subsection{DoA Estimation Performance}
\begin{figure}[!t]
	\centering
	\includegraphics[width=0.78\linewidth]{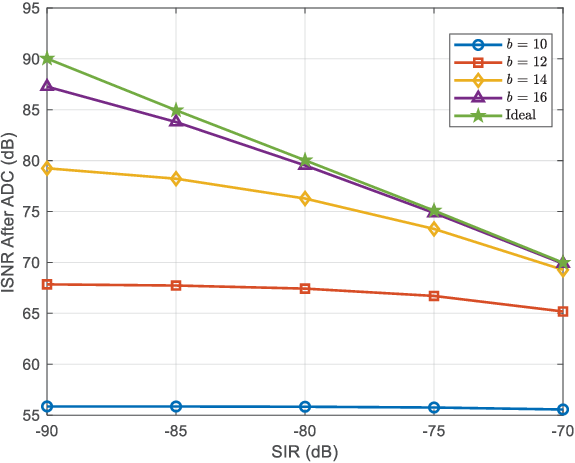}
	\caption{Effective ISNR at the ADC output for different ADC resolutions.}
	\label{SIR after ADC}
\end{figure}

Figure \ref{RMSE of jamming signal with different quantization bits} evaluates the Root Mean Square Error (RMSE) of the jamming DoA estimation versus the input SIR for various ADC resolutions. The results reveal distinct behaviors based on the quantization level. For low-resolution ADCs (e.g., 10-bit and 12-bit), the RMSE remains stable and low, exhibiting a performance floor even as the input SIR varies from -90 dB to -70 dB. This phenomenon is directly explained by Fig. \ref{SIR after ADC}, which shows the interference-to-signal-plus-noise ratio (ISNR) after the ADC quantization stage. Despite the wide variation in input SIR, the effective ISNR at the ADC output is clamped within a narrow range around the critical threshold (56 dB for the 10-bit case), making the estimation accuracy largely independent of the external SIR. In contrast, for higher-resolution ADCs (14-bit and 16-bit) and the ideal case without quantization, the performance is no longer dominated by the quantization level in this SIR range. Consequently, their RMSE curves in Fig. \ref{RMSE of jamming signal with different quantization bits} show a clear upward trend, indicating that estimation accuracy degrades as the input SIR increases (i.e., the jamming becomes weaker).

\begin{figure}[!t]
	\centering
	\includegraphics[width=0.775\linewidth]{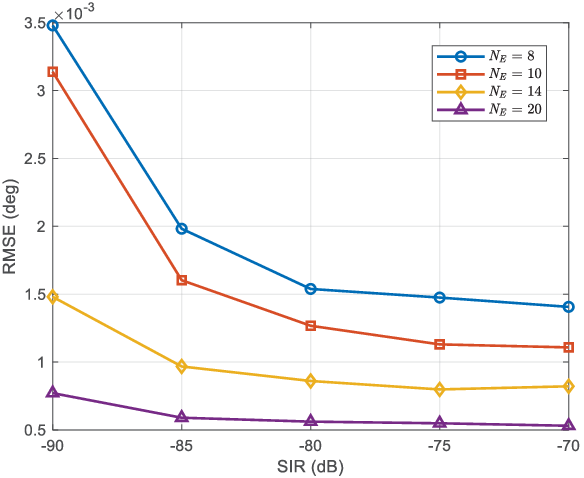}
	\caption{RMSE of desired signal DoA estimation for different $N_E$.}
	\label{RMSE of desired signal with different numbers of antennas}
\end{figure}

After the initial analog suppression stage significantly reduces the jamming power, the desired signal can pass through the ADCs without being blocked, enabling the estimation of its DoAs. Fig. \ref{RMSE of desired signal with different numbers of antennas} presents the RMSE for the desired signal DoA estimation versus the input SIR for different $N_E$. The results show a clear trend where the RMSE of the signal DoA estimation decreases as the input SIR increases, demonstrating the effectiveness of the preceding suppression stage. By creating a more favorable effective SIR at the ADC input, it establishes a viable condition for subsequent signal processing. As with jamming estimation, increasing $N_E$ leads to a lower RMSE for the desired signal, confirming the benefits of a larger antenna aperture.

\begin{figure}[!t]
	\centering
	\includegraphics[width=0.789\linewidth]{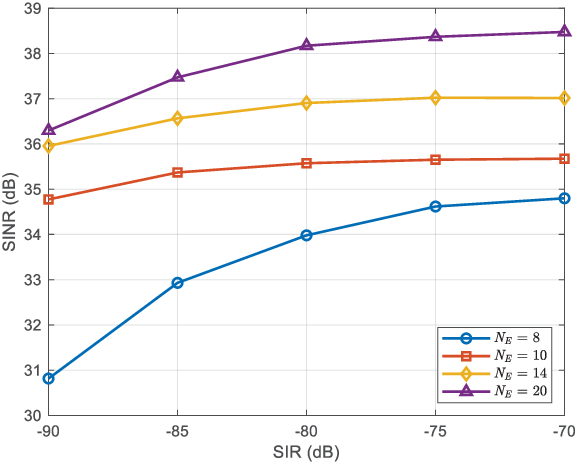}
	\caption{Output SINR after jamming suppression for different $N_E$.}
	\label{Final sinr with different numbers of antennas}
\end{figure}

\subsection{Hybrid Beamforming and Overall Performance Improvement} 
With accurate DoA estimates for both the desired signal and the jamming, the final stage involves jointly optimizing the analog DARISAA phases $\bf{V}$ and the digital beamformer $\bf{w}$ to maximize the output SINR.

 The final SINR at the receiver output is shown in Fig. \ref{Final sinr with different numbers of antennas}. The proposed scheme achieves a significant SINR that is positively correlated with the number of elements per DARISA, $N_E$. This performance enhancement stems from the larger physical aperture, which provides two key benefits: more degrees of freedom for the joint beamforming optimization and more accurate DoA estimates for both jamming and desired signals.  This result highlights the combined power of precise analog nulling of jamming and coherent combining of analog and digital domains. Also, it demonstrates that this system not only overcomes ADC saturation but also achieves high-performance matched reception in severely challenging jamming environments.
 
\begin{figure}[!t]
	\centering
	\includegraphics[width=0.8\linewidth]{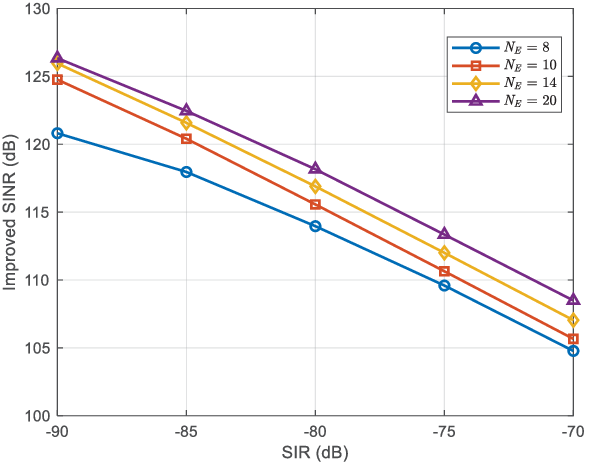}
	\caption{Improved SINR for different $N_E$.}
	\label{Improved sinr with different numbers of antennas}
\end{figure}

Figure \ref{Improved sinr with different numbers of antennas} presents the improved SINR, defined as the difference between the final output SINR and the initial input SINR (both in dB). The figure reveals that the SINR gain decreases as the input SIR increases. This trend does not indicate a performance degradation, but rather highlights the effectiveness of the system. The greatest potential for improvement exists in the most severely jammed scenarios (i.e., lowest input SIR). In these cases, the system's ability to suppress strong jamming levels translates into significant matching gains. As the input jamming becomes weaker, the output SINR still improves, as seen in Fig. \ref{Final sinr with different numbers of antennas}, but the relative gain is smaller because the initial condition is less severe.

\begin{figure}[!t]
	\centering
	\includegraphics[width=0.8\linewidth]{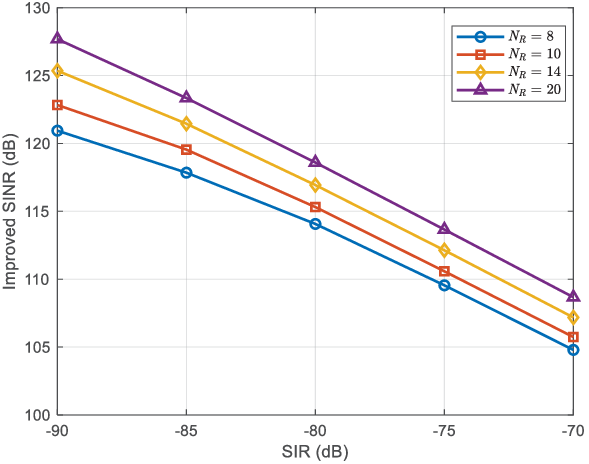}
	\caption{Improved SINR for different $N_R$.}
	\label{Improved sinr with different numbers of RFs}
\end{figure}

To investigate the hardware trade-off between the number of RF chains $N_R$ and the number of elements per DARISA $N_E$, Fig. \ref{Improved sinr with different numbers of RFs} shows the improved SINR for different numbers of RF chains while keeping $N_E$ constant. The similar performance curves suggest that reducing the number of expensive RF chains can be effectively compensated by increasing the number of low-cost metasurface elements in each DARISA. This demonstrates a key advantage of the proposed system, which is the potential to design more cost-effective and energy-efficient receivers without significant performance degradation.


\begin{figure}[!t]
	\centering
	\includegraphics[width=0.8\linewidth]{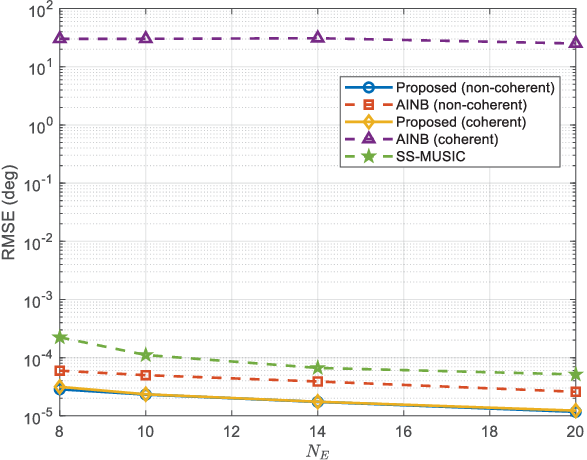}
	\caption{RMSE of jamming DoA estimation with different schemes.}
	\label{RMSE of jamming signal with different schemes}
\end{figure}

\begin{figure}[!t]
	\centering
	\includegraphics[width=0.8\linewidth]{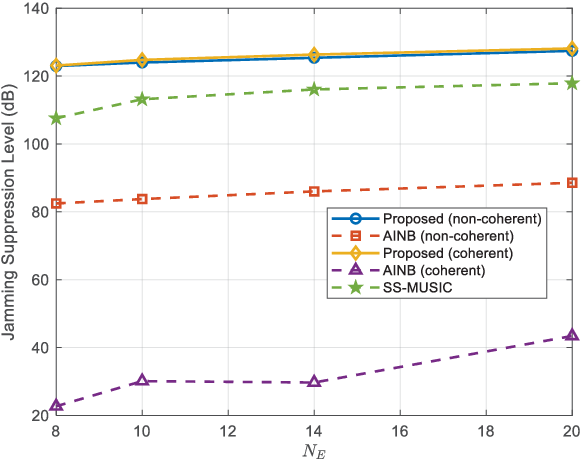}
	\caption{Analog jamming suppression level with different schemes.}
	\label{Jamming suppression level with different schemes}
\end{figure}

Finally, crucial comparisons involving the proposed scheme and a conventional AINB scheme \cite{ref5} under both coherent and non-coherent signal conditions, as well as the spatial smoothing Root-MUSIC (SS-MUSIC) \cite{18}, are provided in Fig. \ref{RMSE of jamming signal with different schemes} and Fig. \ref{Jamming suppression level with different schemes}. For a fair comparison, the SS-MUSIC scheme is integrated into our three-stage architecture to replace the proposed TSS method while retaining the identical analog pre-suppression mechanism. As shown in Fig. \ref{RMSE of jamming signal with different schemes}, the AINB scheme completely fails to estimate the DoAs of coherent jamming signals, resulting in an extremely high RMSE. While the SS-MUSIC scheme successfully handles coherent signals, our proposed TSS method significantly outperforms it by achieving much higher DoA estimation accuracy. This gap arises because SS-MUSIC incurs severe aperture loss by dividing the array into overlapping subarrays to restore signal rank, whereas our TSS method decorrelates signals temporally to fully preserve the spatial degrees of freedom. Also, an important observation from this figure is that as $N_E$ increases, the RMSE of the proposed scheme consistently decreases. This is because a larger number of metasurface elements provides a larger effective aperture, leading to higher spatial resolution and more accurate DoA estimation.

Furthermore, the analog jamming suppression levels of the different schemes are provided in Fig. \ref{Jamming suppression level with different schemes}. The proposed scheme achieves over 120 dB of suppression for coherent signals, whereas the AINB scheme provides negligible suppression. Although the SS-MUSIC achieves considerable suppression by resolving the coherent paths, its performance remains noticeably inferior to the proposed scheme. This is because its inherent aperture loss compromises DoA accuracy and subsequently degrades spatial nulling precision. Additionally, it can be seen that the suppression level consistently improves as $N_E$ increases, which is attributable to the enhanced degrees of freedom for null steering and more accurate jamming DoA estimation. Crucially, the performance of our scheme is nearly identical for both coherent and non-coherent cases, validating the effectiveness of TSS technique in signal decorrelation. These results clearly demonstrate the robustness and superiority of our scheme, especially in realistic scenarios involving severe coherent multipath jamming.

\section{Conclusion}\label{8}
This paper addresses the issue of AGC-limited ADC blocking caused by a low SIR in suppressive jamming scenarios, where the weak desired signal becomes unresolvable after gain compression and finite-resolution quantization.
A novel jamming suppression and matching reception scheme for finite-precision ADCs is proposed, leveraging the reconfigurable characteristics of DARISAA. The scheme enables fine-grained estimation of channel multipath information, which is essential for effective jamming mitigation. Based on this, a two-stage DARISAA spatial filtering and ADC-aware reception design is introduced. 
This approach effectively mitigates AGC-induced quantization blocking caused by jamming, ensuring that the desired signal successfully passes through to the digital domain for signal processing. Simulation results demonstrate that the proposed scheme not only achieves excellent jamming suppression but also significantly improves the SINR, effectively enabling the matching reception of desired signals even in the presence of suppressive jamming.

\vfill


\begin{thebibliography}{99}
	\bibitem{ref1}
	X. Qi, M. Peng, H. Zhang and X. Kong, ``Anti-jamming hybrid beamforming design for millimeter-wave massive MIMO systems,'' \emph {IEEE Trans. Wireless Commun.}, vol. 23, no. 8, pp. 9160-9172, Aug. 2024.
	
	\bibitem{ref2}
	Z. Chu et al., ``Throughput improvement for RIS-empowered wireless powered anti-jamming communication networks (WPAJCN),'' \emph {IEEE Trans. Inf. Forensic Secur.}, vol. 20, pp. 4622-4637, 2025.
	
	\bibitem{ref3}
	J. Liu, G. Yang, Y.-C. Liang, and C. Yuen, ``Max-min fairness in RIS-assisted anti-jamming communications: Optimization versus deep reinforcement learning approaches,'' \emph {IEEE Trans. Commun.}, vol. 72, no. 7, pp. 4476–4492, July 2024.
	
	\bibitem{ref4}
	Y. Sun et al., ``Active-passive cascaded RIS-aided receiver design for jamming nulling and signal enhancing,'' \emph {IEEE Trans. Wireless Commun.}, vol. 23, no. 6, pp. 5345–5362, June 2024.
	
	\bibitem{addref1}
	R. Han, T. Zhang, H. Zhong and Y. Wang, ``A dynamic UAVs cooperative suppressive jamming method with joint task assignment and bandwidth allocation,''  \emph {IEEE Trans. Aerosp. Electron. Syst.}, vol. 61, no. 5, pp. 11469-11488, Oct. 2025.
	
	\bibitem{13}
	W. Zhang, Y. Jiang, B. Zhou and D. Hu, ``Hybrid interference mitigation using analog prewhitening,'' \emph{IEEE Trans. Wireless Commun.}, vol. 20, no. 10, pp. 6595-6605, Oct. 2021.
	
	\bibitem{3}
	A. Singh and M. K. Mandal, ``Electronically tunable reflection type 
	phase shifters,'' \emph{IEEE Trans. Circuits Syst. II-Express Briefs},  vol. 67, no. 3, pp. 425-429, Mar. 2020.
		
	\bibitem{4}
	A. M. Abbosh, ``Compact tunable reflection phase shifters using short 
	section of coupled lines,'' \emph{IEEE Trans. Microw. Theory Tech.},  vol. 60, no. 8, pp. 2465-2472, Aug. 2012.

	\bibitem{5}
	R. W. Irazoqui and C. J. Fulton, ``Spatial interference nulling before RF frontend for fully digital phased arrays,'' \emph{ IEEE Access}, vol. 7, pp. 151261-151272, 2019.
	
	\bibitem{6}
	Y. R. Guo, M. T. Lin, Z. F. Wu, et al., ``RF front-end multi-channel amplitude and phase weighting loop for interference mitigation,'' \emph{Chin. J. Radio Sci.}, vol. 38, no. 5, pp. 1-8, 2023.

	\bibitem{7}
	R. J. Vaccaro and B. F. Harrison, ``Optimal matrix-filter design,'' \emph{IEEE Trans. Signal Process.}, vol. 44, no. 3, pp. 705-709, Mar. 1996.
	
	\bibitem{8}
	J. H. C. van den Heuvel, J. -P. M. G. Linnartz, P. G. M. Baltus and D. Cabric, ``Full MIMO spatial filtering approach for dynamic range reduction in wideband cognitive radios,'' \emph{ IEEE Trans. Circuits Syst. I-Regul. Pap.}, vol. 59, no. 11, pp. 2761-2773, Nov. 2012.
	
	\bibitem{9}
	V. Venkateswaran and A. -J. van der Veen, ``Analog beamforming in MIMO communications with phase shift networks and online channel estimation,'' \emph {IEEE Trans. Signal Process.}, vol. 58, no. 8, pp. 4131-4143, Aug. 2010.
	
	\bibitem{10}
	M. A. Alaei, S. Golabighezelahmad, P. -T. de Boer, F. E. van Vliet, E. A. M. Klumperink and A. B. J. Kokkeler, ``Interference mitigation by adaptive analog spatial filtering for MIMO receivers,'' \emph { IEEE Trans. Microw. Theory Tech.}, vol. 69, no. 9, pp. 4169-4179, Sept. 2021.
	
	\bibitem{11}
	W. Zhang, X. Xia, Y. Fu and X. Bao,  ``Hybrid and full-digital beamforming in mmWave massive MIMO systems: A comparison considering low-resolution ADCs,'' \emph{China Commun.}, vol. 16, no. 6, pp. 91-102, June 2019.
		
	\bibitem{ref5}
	K. Wu, J. A. Zhang, X. Huang, Y. J. Guo, D. N. Nguyen, A. Kekirigoda, and K.-P. Hui, ``Analog-domain suppression of strong interference using hybrid antenna array,'' \emph{Sensors}, vol. 22, no. 6, art. no. 2417, Mar. 2022. 
	
	\bibitem{ref11}
	J. Bai, H. -M. Wang and L. Jin, ``Dynamic agile reconfigurable intelligent surface antenna (DARISA) MIMO: DoF analysis and effective DoF optimization,'' \emph{IEEE Trans. Wireless Commun.}, vol. 25, pp. 2197-2212, 2026.

	\bibitem{ref12}
	X. Wei and H.-M. Wang, ``Multi-user downlink with reconfigurable intelligent metasurface antennas (RIMSA) array,'' \emph{IEEE Trans. Veh. Technol.}, vol. 74, no. 12, pp. 18914-18929, Dec. 2025.
	
	\bibitem{addref12}
	Y. Huang, H. -M. Wang, Q. Yan and Z. Wang, ``LLM-RIMSA: Large language models driven reconfigurable intelligent metasurface antenna systems,'' \emph{IEEE J. Sel. Areas Commun.}, vol. 44, pp. 2479-2493, 2026.
	
	\bibitem{ref13}
	Z. Wang, Y. Huang, W. Liu, and H.-M. Wang, ``Anti-jamming sensing with distributed reconfigurable intelligent metasurface antennas,'' \emph{IEEE Trans. Wireless Commun.}, vol. 25, pp. 6681-6694, 2026.


	\bibitem{Wu2025}
	Q. Wu et al., ``Intelligent reflecting surfaces for wireless networks: Deployment architectures, key solutions, and field trials,'' \emph{IEEE Wireless Commun.}, vol. 32, no. 6, pp. 141-148, Dec. 2025.

	\bibitem{DongWang2022}
	L. Dong, H. -M. Wang and J. Bai, ``Active reconfigurable intelligent surface aided secure transmission,'' \emph{IEEE Trans. Veh. Technol.}, vol. 71, no. 2, pp. 2181-2186, Feb. 2022.
	
	\bibitem{12}
	O. Orhan, E. Erkip and S. Rangan, ``Low power analog-to-digital conversion in millimeter wave systems: Impact of resolution and bandwidth on performance,'' in \emph{Proc. Inf. Theory Appl. Workshop (ITA)}, San Diego, CA, USA, Oct. 2015, pp. 191-198.
	
	\bibitem{14}
	L. Fan, S. Jin, C. -K. Wen and H. Zhang, ``Uplink achievable rate for massive MIMO systems with low-resolution ADC,'' \emph{IEEE Commun. Lett.}, vol. 19, no. 12, pp. 2186-2189, Dec. 2015.
	
	\bibitem{GershoGray1992}
	A. Gersho and R. M. Gray, \emph{Vector Quantization and Signal Compression}. Boston, MA, USA: Kluwer Academic Publishers, 1992.
	
	
	\bibitem{15}
	R. Schmidt, ``Multiple emitter location and signal parameter estimation,'' \emph{IEEE Trans. Antennas Propag.}, vol. 34, no. 3, pp. 276-280, Mar. 1986.
	
	\bibitem{16}
	R. Roy and T. Kailath, ``ESPRIT-estimation of signal parameters via rotational invariance techniques,'' \emph{IEEE Trans. Acoust., Speech, Signal Process.}, vol. 37, no. 7, pp. 984-995, July 1989.

	\bibitem{17}
	Tie-Jun Shan, M. Wax and T. Kailath, ``On spatial smoothing for direction-of-arrival estimation of coherent signals,'' \emph{IEEE Trans. Acoust., Speech, Signal Process.}, vol. 33, no. 4, pp. 806-811, Aug. 1985.
	
	\bibitem{18}
	S. E. Sorkhabi and K. Rambabu, ``Multi-target DoA estimation with mmWave MIMO radar using limited number of sensors,''  \emph{ IEEE Trans. Veh. Technol.}, vol. 74, no. 9, pp. 13783-13794, Sept. 2025.
	
	\bibitem{19}
	J. Pan, M. Sun, Y. Wang and X. Zhang, ``An enhanced spatial smoothing technique with ESPRIT algorithm for direction of arrival estimation in coherent scenarios,'' \emph{IEEE Trans. Signal Process.}, vol. 68, pp. 3635-3643, 2020.
	
	\bibitem{20}
	F.-M. Han and X.-D. Zhang, ``An ESPRIT-like algorithm for coherent DOA estimation,'' \emph{IEEE Antennas Wirel. Propag. Lett.}, vol. 4, pp. 443-446, 2005.
	
	\bibitem{21}
	W. Zhang, Y. Han, M. Jin and X. -S. Li, ``An improved ESPRIT-like algorithm for coherent signals DOA estimation,'' \emph{IEEE Commun. Lett.}, vol. 24, no. 2, pp. 339-343, Feb. 2020.
	
	
	\bibitem{WangLiu2023}
	Q. Wang, L. Liu, and S. Zhang, ``MUSIC algorithm for IRS-assisted AOA estimation,'' in \emph{Proc. IEEE 98th Veh. Technol. Conf. (VTC-Fall)}, Hong Kong, China, Oct. 2023, pp. 1–5.
	
	\bibitem{22}
	A. J. Barabell, ``Improving the resolution performance of eigenstructure-based direction-finding algorithms,'' in \emph{Proc. IEEE Int. Conf. Acoust., Speech, Signal Process. (ICASSP)}, Boston, MA, USA, Apr. 1983, pp. 336-339.

	\bibitem{23}
	P.-A. Absil, R. Mahony, and R. Sepulchre, \emph{Optimization Algorithms on Matrix Manifolds}. Princeton, NJ, USA: Princeton Univ. Press, 2009.
	
	\bibitem{ref24}
	D. P. Bertsekas, \emph{Constrained Optimization and Lagrange Multiplier Methods}. New York: Academic, 1982.
	
	\bibitem{ref25}
	J. Nocedal and S. J. Wright, \emph{Numerical Optimization}, 2nd ed. New York, NY, USA: Springer, 2006.
		
	\bibitem{add-r3}
	W. Xiong et al., ``Cooperative double IRS aided secure communication for MIMO-OFDM systems,'' \emph{IEEE Trans. Veh. Technol.}, vol. 75, no. 6, pp. 11943-11948, June 2026.
	
	\bibitem{24}
	G. H. Golub and C. F. Van Loan, \emph{Matrix Computations}, 4th ed. Baltimore, MD, USA: Johns Hopkins Univ. Press, 2013.
	
\end{thebibliography}
\end{document}